\documentclass[review]{elsarticle}

\usepackage{lineno}
\usepackage{graphics}
\usepackage{graphicx}
\usepackage{color}
\usepackage{subfigure}
\usepackage{multicol}
\usepackage{multirow}
\usepackage{amssymb}
\usepackage{amsfonts}
\usepackage{float}
\usepackage{booktabs}
\usepackage{psfig}
\usepackage{amsmath}
\usepackage{amssymb}
\usepackage{mathrsfs}
\usepackage{algpseudocode}
\usepackage{epstopdf}
\usepackage{bm}
\usepackage{booktabs}
\usepackage{hyperref}
\usepackage{caption}
\biboptions{numbers,sort&compress}
\modulolinenumbers[5]

\journal{Pattern Recognition}

\begin{document}

\begin{frontmatter}

\title{Knowledge-aware Deep Framework for Collaborative Skin Lesion Segmentation and Melanoma Recognition}

\author[ntu]{Xiaohong Wang}
\ead{E150023@e.ntu.edu.sg}

\author[ntu]{Xudong Jiang\corref{cor1}}
\ead{exdjiang@ntu.edu.sg}

\author[ntu]{Henghui Ding}
\ead{ding0093@ntu.edu.sg}

\author[csu]{Yuqian Zhao}
\ead{zyq@csu.edu.cn}

\author[sutd]{Jun Liu\corref{cor1}}
\ead{junliu@sutd.edu.sg}

\cortext[cor1]{Corresponding author}
\address[ntu]{School of Electrical and Electronic Engineering, Nanyang Technological University, Singapore, 639798}
\address[csu]{School of Automation, Central South University, Hunan, 410083}
\address[sutd]{Information Systems Technology and Design Pillar, Singapore University of Technology and Design, Singapore, 487372}

\begin{abstract}
Deep learning techniques have shown their superior performance in dermatologist clinical inspection.
Nevertheless, melanoma diagnosis is still a challenging task due to the difficulty of incorporating the useful dermatologist clinical knowledge into the learning process. In this paper, we propose a novel knowledge-aware deep framework that incorporates some clinical knowledge into collaborative learning of two important melanoma diagnosis tasks, i.e., skin lesion segmentation and melanoma recognition.
%
Specifically, 
to exploit the knowledge of morphological expressions of the lesion region and also the periphery region  for melanoma identification, a lesion-based pooling and shape extraction (LPSE) scheme is designed, which transfers the structure information obtained from skin lesion segmentation into melanoma recognition.
Meanwhile, to pass the skin lesion diagnosis knowledge from melanoma recognition to skin lesion segmentation,
an effective diagnosis guided feature fusion (DGFF) strategy is designed.
Moreover,
we propose a recursive mutual learning mechanism that further promotes the inter-task
cooperation, and thus iteratively improves the joint learning capability of the model for both skin lesion segmentation and melanoma recognition.
Experimental results on two publicly available skin lesion datasets show the effectiveness of the proposed method for melanoma analysis.
\end{abstract}


\end{frontmatter}


\section{Introduction}

Melanoma is one of the most malignant skin cancer that increases rapidly throughout the world \cite{kawahara2018seven,hu2018deep,yuan2017automatic,barata2020explainable}. Timely treatment of the melanoma can efficiently improve the survive rate of the patients. 
Dermoscopic images captured by digital imaging devices, offer the magnified visualization of the melanoma, and thus assist the dermatologists in examining the melanoma based on a set of complex visual characteristic of the lesion. Computer-aided diagnosis (CAD) system provides an effective way that allows dermatologists' clinical inspection of the skin lesion in dermoscopic images. A CAD system for melanoma analysis generally contains two crucial functions: lesion segmentation and melanoma recognition. Specifically, the task of segmentation~\cite{shuai2019toward,ding2019semantic,ding2020phraseclick,ding2021vision,ding2021interaction,liu2021few,liuweid2021few, zhang2021prototypical} aims to divide a dermoscopic image into the skin lesion parts and background parts, i.e., it is a pixel-wise classification process to generate more conceptual saliency information for melanoma analysis \cite{zortea2017simple, schmid1999segmentation,zhou2009anisotropic,zhou2011gradient,ma2016novel}. 
Meanwhile, melanoma recognition is an image-level classification task that aims to identify the skin lesion types, such as melanoma, seborrheic keratosis, and benign nevi \cite{bi2020multi,sadeghi2013detection,saez2014model,yu2018melanoma,yang2018classification}.

\begin{figure}[t]
\hspace*{-0.3cm}
\vspace*{-0.3cm}
\centering
\includegraphics[width=10cm,height=7cm]{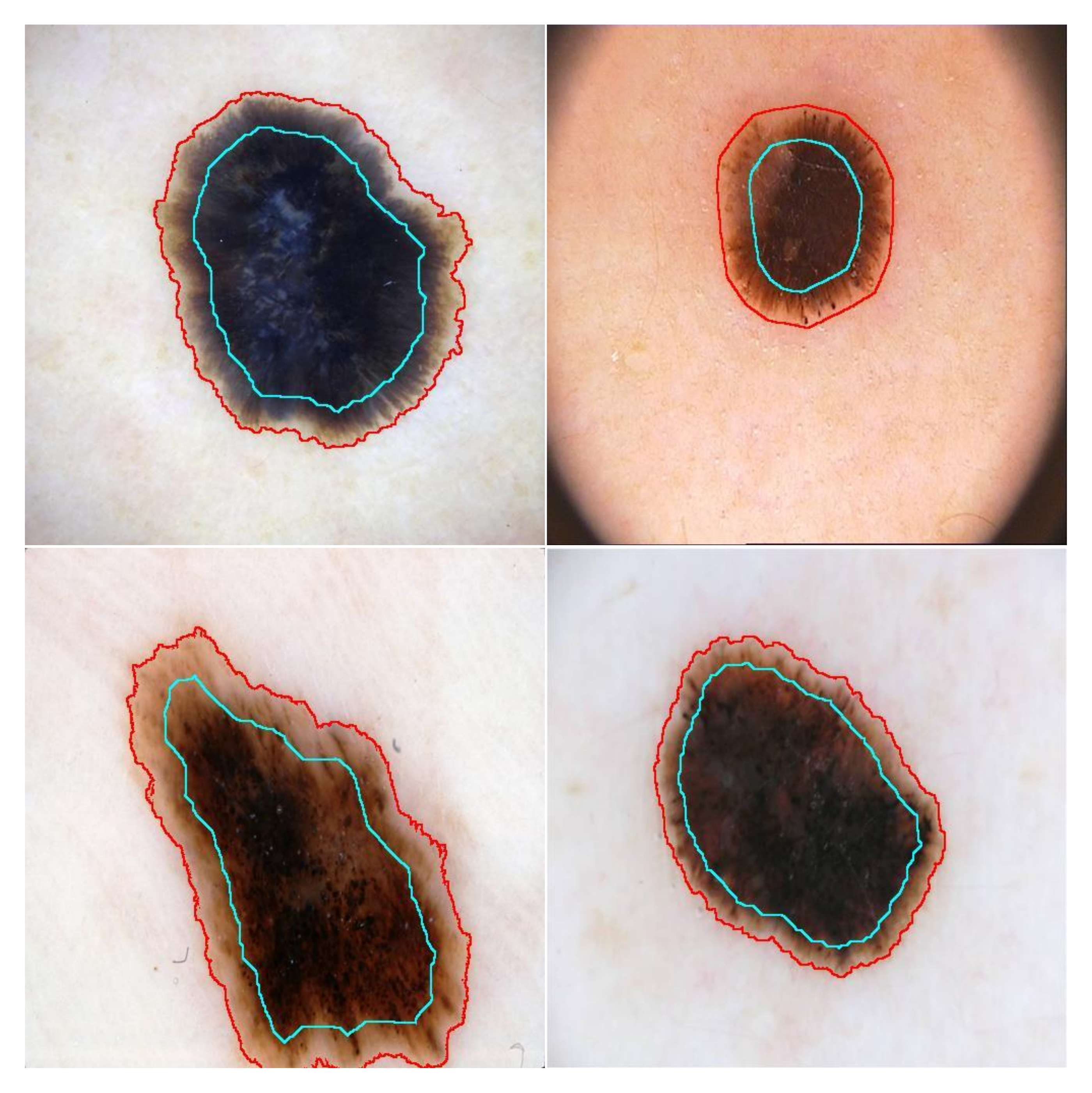}
\caption{Visualization examples of skin lesions. Red and blue contours indicate the periphery and center regions of the skin lesions respectively. Some important morphological expressions for recognizing skin lesions, like streaks, can be observed at the periphery regions.}
\label{fig:1}
\end{figure}

Recent deep learning-based methods~\cite{cai2021unified,sun2021m2iosr,wang2021adaptive,li2021else,tong2021directed,li2021mine,mei2020object} have shown their promising achievements in both lesion segmentation~\cite{bi2017dermoscopic,yuan2017improving,wang2019dermoscopic,zhang2019dsm,singh2019fca} and melanoma recognition~\cite{ harangi2018skin,yu2018melanoma,mahbod2019fusing,liang2019multi,gessert2019skin,sultana2018deep,hagerty2019deep,zhang2019attention,yang2018classification,yu2017automated}. Existing approaches usually train the task-specific models to perform skin lesion segmentation and melanoma recognition separately, and do not explicitly consider the dermoscopists' clinical criteria for melanoma inspection. For example, most of the existing methods 
\cite{yu2018melanoma,harangi2018skin,mahbod2019fusing,liang2019multi,gessert2019skin,sultana2018deep,hagerty2019deep,zhang2019attention,yang2018classification,yu2017automated} heavily depend on the abstract feature representation obtained at the high layer of the network, and they often leverage the global information of the whole feature maps for melanoma identification, i.e., melanoma recognition is often conducted in a similar fashion as other general object classification tasks \cite{he2016deep,he2015spatial}. However,
in the domain of melanoma inspection, such a fashion has a high potential of losing important pathological patterns and morphology knowledge that are crucial for melanoma recognition. Generally, for dermatologists, they identify melanoma via certain prior criteria, like the statistical information of the lesion color's variation and the border's irregularity \cite{gonzalez2018dermaknet,hagerty2019deep,xie2016melanoma}. Moreover, clinical studies \cite{sadeghi2013detection,tajeddin2018melanoma,riaz2018active} have  confirmed that the characteristics of the peripheral lesion 
are very useful morphological expressions for identifying melanoma, as shown in Fig. \ref{fig:1}. However, these types of knowledge have not been explicitly considered yet by current deep learning-based methods \cite{mahbod2019fusing,liang2019multi,gessert2019skin,sultana2018deep,hagerty2019deep,zhang2019attention,yu2017automated}.

In this paper, we propose a novel knowledge-aware framework that is able to exploit the clinical knowledge within the deep feature learning process for melanoma analysis. Specifically, in our framework, both the skin lesion segmentation task and the melanoma recognition task are learned and performed via a joint deep network, such that the clinical knowledge can be exploited and transferred with the mutual guidance and assistance of these two tasks.

Concretely, to promote the performance of the deep learning framework on melanoma recognition,
we propose to incorporate the clinical knowledge by explicitly considering the morphological expression of the lesion area and also the periphery region.
To achieve this, we design a novel lesion-based pooling and shape extraction (LPSE) module that transfers the lesion structure information obtained from the skin lesion segmentation task to the melanoma recognition task, and thus we are able to embed the morphological operation into our deep network. With the integration of morphological analysis of the skin lesion structure, our network is thus able to selectively learn the informative features containing useful statistical information from both the lesion center region and also the border region. 
Compared with the features produced by direct global average pooling in most of the existing deep learning-based methods \cite{harangi2018skin,mahbod2019fusing,liu2021towards,liang2019multi,mei2019deepdeblur,liu2019feature,gessert2019skin,sultana2018deep,hagerty2019deep,zhang2019attention}, our network generates more discriminative lesion representation for melanoma recognition. 

Melanoma and non-melanoma lesions generally have very different pathological feature representations (e.g., more border irregularity and inhomogeneous textures for melanoma lesions). With lesion class information, the segmentation network can generate more discriminative feature representation for detecting different types of skin lesions from dermoscopic images. To improve skin lesion segmentation performance for both the melanoma and non-melanoma classes, we propose a new unit called diagnosis guided feature fusion (DGFF) that incorporates the lesion diagnosis information from melanoma recognition task into skin lesion segmentation task. With the guidance of skin lesion class information learned from the melanoma recognition task, our DGFF achieves more discriminative feature representation for each lesion class and thus enhances the capability of our network in segmenting the skin lesion regions from the dermoscopic images.

Moreover, to seamlessly achieve both the aforementioned skin lesion segmentation task and the melanoma recognition task within our end-to-end network, we also propose a recursive mutual learning scheme that enables effective inter-task cooperation and recurrently improves the joint leaning capability of the model on both tasks. The proposed recursive mutual learning scheme systematically exploits the mutual guidance signals generated between skin lesion segmentation and melanoma recognition, and thus simultaneously boosts the performance of both tasks. The main contributions of this paper are summarized as follows:

$\bullet$ We propose a novel end-to-end deep framework that is able to perform skin lesion segmentation and melanoma recognition jointly,  where the  clinical  knowledge  is  exploited  and  transferred  with  the  mutual  guidance between these two tasks.

$\bullet$ We design a lesion-based pooling and shape extraction module to transfer the lesion structure information from the skin lesion segmentation task to the melanoma recognition task, by embedding the morphological analysis into the feature learning process of our network, which helps generate informative clinically interested features for melanoma recognition.

$\bullet$ We propose a diagnosis guided feature fusion scheme to pass the lesion class information from the melanoma recognition task into the skin lesion segmentation task, which generates discriminative representations for different types of skin lesions.

$\bullet$ We design a recursive mutual learning method that further enhances the joint learning ability of the proposed model for both skin lesion segmentation and melanoma recognition.


\section{Related Work}
Deep learning-based methods have been used for melanoma diagnosis during recent years. Skin lesion segmentation and melanoma recognition are both important research tasks for melanoma inspection. In this section, we review the deep learning-based methods that address these two tasks.

\textbf{Skin lesion segmentation:} The work of \cite{yuan2017automatic} introduced a deep fully convolutional network with Jaccard distance for skin lesion segmentation. Bi \emph {et al}.~\cite{bi2017dermoscopic} refined the skin lesion segmentation performance by integrating the multiple embedded FCN stages. Yuan \emph {et al}.~\cite{yuan2017improving} extended their previous work~\cite{yuan2017automatic} by developing a deeper network framework with a smaller kernels. Al-masni \emph {et al}.~\cite{al2018skin} proposed a full resolution convolutional networks (FrCn) to generate the full resolution features of each pixel of the input dermoscopic images. Li \emph {et al}.~\cite{li2018dense} detected skin lesion with a dense deconvolutional network (DDN). Tu \emph {et al}.~\cite{tu2019dense} achieved skin lesion segmentation by exploiting a dense-residual network with adversarial learning. Bi \emph {et al}.~\cite{bi2019step} used a deep class-specific learning framework to learn the important visual characteristics of the skin lesions of each individual class (melanoma vs. non-melanoma). Tang \emph{et al}.~\cite{tang2019efficient} introduced a skin lesion segmentation approach based on the separable-Unet with stochastic weight averaging. The authors in~\cite{nasr2019dense} applied a fully convolutional network with dense pooling layers to delineate the boundary of the skin lesion. Wei \emph {et al}.~\cite{wei2019attention} adopted an attention-based denseUnet network with adversarial training for skin lesion segmentation. The authors in~\cite{wang2019bi} employed a skin lesion segmentation framework with bi-directional dermoscopic feature learning and multi-scale consistent decision fusion. Recent multi-path-based deep learning architectures are also used in \cite{zhang2019dsm,singh2019fca} for skin lesion segmentation.

\begin{figure}[t]
\hspace*{-2.5cm}
\includegraphics[width=17 cm,height=6cm ]{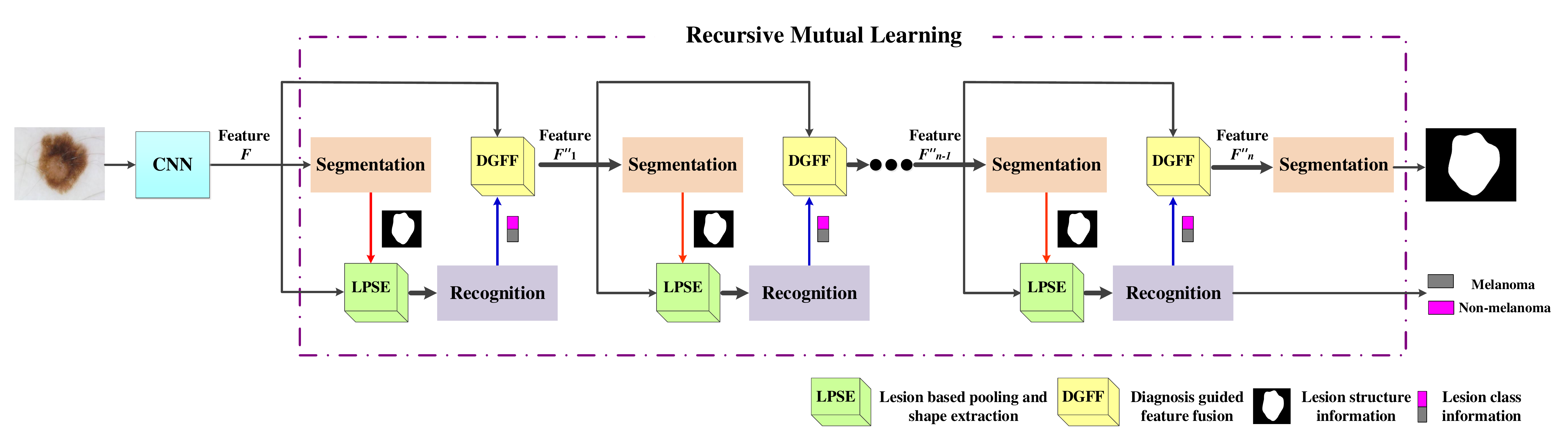}
\caption{The flowchart of the proposed knowledge-aware deep framework for skin lesion segmentation and melanoma recognition. Our proposed knowledge-aware deep framework collaboratively achieves the performance enhancement of skin lesion segmentation and melanoma recognition by recursively utilizing mutual benefits from each individual task, i.e., exploiting lesion structure information from skin lesion segmentation task for recognition, and involving lesion class information from melanoma recognition task for segmentation.}
\label{fig:2}
\end{figure}

\textbf{Melanoma recognition:} Esteva \emph{et al}.~\cite{esteva2017dermatologist} used GoogleNet Inception v3 CNN architecture for melanoma recognition. Harangi~\cite{harangi2018skin} achieved melanoma recognition by fusing the outputs of the classification layers of four different deep neural network architectures. Yu \emph{et al}.~\cite{yu2018melanoma} aggregated local convolutional features extracted from a deep residual network by Fisher vector (FV) encoding strategy. The authors in~\cite{mahbod2019fusing} fused deep features from multiple pre-trained and fine-tuned DNNs at abstraction levels. Liang \emph{et al}.~\cite{liang2019multi} explored a multi-pooling attention learning for skin lesions classification. Gessert \emph{et al}.~\cite{gessert2019skin} built a skin lesion classification framework using CNNs with patch-based attention and diagnosis-guided loss weighting. Sultana \emph{et al}.~\cite{sultana2018deep} applied a deep residual network with regularised fisher framework for the melanoma recognition. Hagerty \emph{et al}.~\cite{hagerty2019deep} combined conventional image processing with deep learning to improve the melanoma diagnosis performance. Zhang \emph{et al}.~\cite{zhang2019attention} exploited an attention residual learning convolutional neural network for skin lesion classification. Gu \emph{et al}.~\cite{gu2019progressive} designed a two-step progressive transfer learning and adversarial domain adaption for skin disease classification. 

Meanwhile, there are also several efforts aiming to improve melanoma recognition by integrating the localization information from skin lesion segmentation task~\cite{yu2017automated,yang2018classification,al2020multiple}. They usually focus on the feature extraction from region of interest and remove the noise signals from background. For example, Yang \emph {et al}.~\cite{yang2018classification} embedded a region average pooling (RAPooling) module into convolutional neural network to reduce the distraction of the background. Yu \emph {et al}.~\cite{yu2017automated} cropped the lesion patch based on the lesion segmentation network, and then applied a deep residual network for melanoma recognition.
Xie \emph {et al}.~\cite{xie2020mutual} designed a deep  framework for sequential lesion segmentation and classification.
Al \emph {et al}.~\cite{al2020multiple} improved the skin lesion disease classification performance by feeding the segmented skin lesions into a convolutional network.


Different from all the aforementioned works, 
in this paper, we propose a novel knowledge-aware deep framework that iteratively exploits the mutual benefits from both the skin lesion segmentation task and the melanoma recognition task, and thus simultaneously enhances the performance of both tasks. By incorporating the clinical knowledge into our deep learning architecture, the reliability of skin lesion analysis is improved. In addition, the proposed multi-stage mutual learning scheme further promotes the inter-task cooperation, and increases the joint learning capability of the model on both skin lesion segmentation and melanoma recognition.

\section{Proposed Network}
In this section, we describe the proposed knowledge-aware deep framework for joint skin lesion segmentation and melanoma recognition in detail. First, we introduce our lesion-based pooling and shape extraction (LPSE) method that transfers the skin lesion structure information from the skin lesion segmentation task to the melanoma recognition task. It is capable of selectively learning the informative clinically interested features from the lesion and its border regions. Second, we propose a diagnosis guided feature fusion (DGFF) scheme that exploits the lesion class information from the melanoma recognition task to boost the pixel-wise classification performance of skin lesion segmentation. Moreover, our design of recursive mutual learning further promotes the collaborative learning ability of the network on both skin lesion segmentation and melanoma recognition. The overall architecture of the proposed knowledge-aware deep framework is shown in Fig. \ref{fig:2}.

\subsection{Lesion-based Pooling and Shape Extraction}

\begin{figure}[t]
\centering
\includegraphics[width=11 cm,height=5cm ]{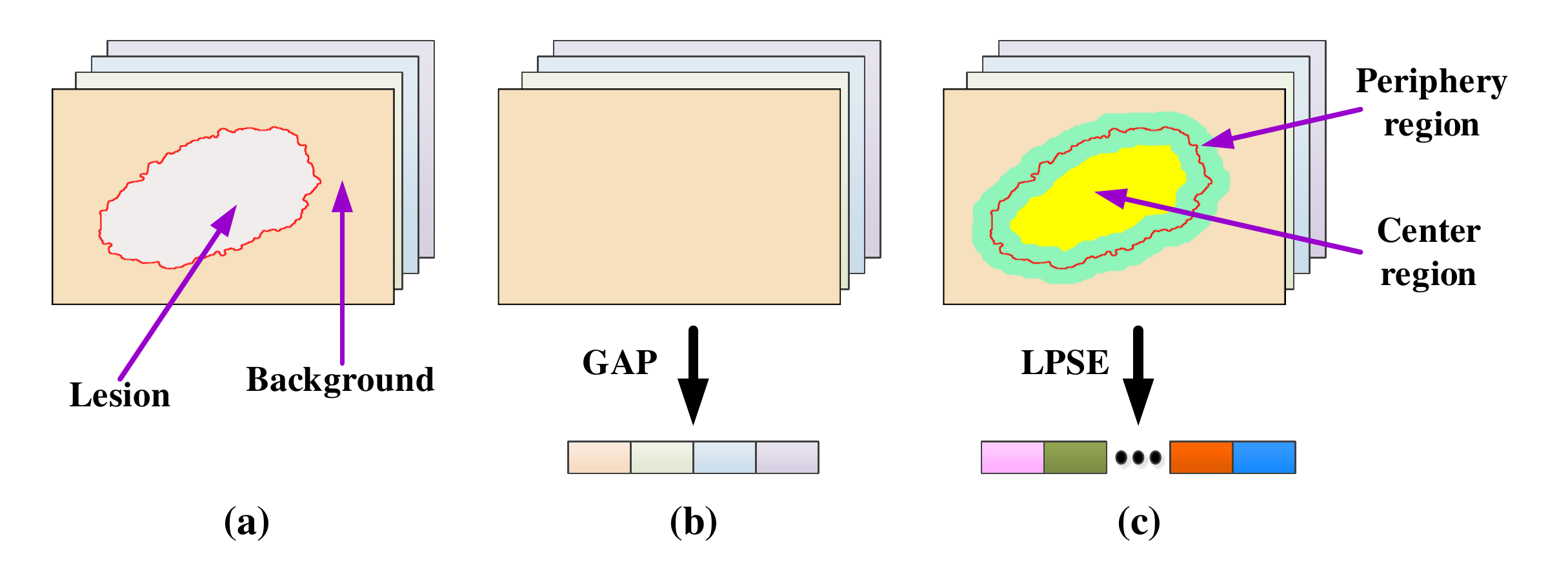}
\vspace*{-0.5cm}
\caption{Example of a network structure with lesion-based pooling and shape extraction. (a) Feature representation for skin lesion and background. (b) Global average pooling. (c) Lesion-based pooling and shape extraction. The red contours indicate the boundary between skin lesion and background.}
\label{fig:3}
\end{figure}

To identify the skin lesion type, dermatologists usually make an inspection by observing the detailed anatomical structures and morphological expressions of the lesion.
Deep learning-based melanoma recognition frameworks can also be used to achieve certain level of identification of the skin lesion as melanoma or non-melanoma, by encoding the convolutional feature information at the high layers. Most of these frameworks use global average pooling (GAP) to generate the global representation for melanoma recognition~\cite{yu2017automated,liang2019multi,mahbod2019fusing,gessert2019skin,zhang2019attention,xie2020mutual}. Fig. \ref{fig:3}(b) shows the process of global average pooling, where the spatial average operation is performed over the entire feature maps of each channel at the last convolutional layer. However, in dermoscopic images, some noises, such as hair, air bubbles, and calibrator, also appear as a part of the dermoscopic image. Thus the information from the noises is also aggregated to the feature representation when directly using global average pooling. Moreover, the domain knowledge (observing the morphological expressions and statistical information of the lesion area, especially the border region) that is often used by dermatologists for melanoma identification, is not specifically considered when using such a global average pooling operation. Therefore, in order to reliably recognize melanoma, it is beneficial to consider the clinical domain knowledge during the feature learning process of the deep network. In this paper, we propose a lesion-based pooling and shape extraction method that explicitly exploits the important clinical knowledge by embedding morphology analysis within our network, which thus generates informative feature representation for melanoma recognition.


Inspired by the process that dermatologists inspect  melanoma, we first aim to delineate a desirable shape of skin lesion. Benefiting from the skin lesion segmentation task that is able to encode the input image into the preliminary representation for lesion parsing, we formulate the lesion shape by exploiting the information from the skin lesion segmentation process. Let $S^{c}_{p}$ denote the score maps generated from the pixel-wise skin lesion segmentation task, i.e.,
{\color{black}
\begin{equation}
S^{c}_{p} = \mathfrak{F}_{a}(F; \theta_{a}),
\end{equation}}
\noindent where $p$ is the spatial position, and $c$ indicates the lesion class or background class. $\mathfrak{F}_{a}$ is the linear classification function, and $\theta_{a}$ denotes the corresponding parameters. {\color{black} $F$ denote the feature maps computed from the layer before the global average pooling layer of the baseline network, as shown in Fig. 2.} 
Thus the lesion existence probability can be formulated as $o^{c}_{p} = \mathfrak{F}_{b}(S^{c}_{p})$, where $\mathfrak{F}_{b}$ is the softmax function. Then we convert the lesion probability map ($o^{c}_{p}$) into a binary mask, $M$, using a fixed threshold $0.5$, such that the skin lesion region and the background area are clearly separated. Concretely, if the probability that a candidate pixel belongs to skin lesion is larger than the threshold 0.5, it is classified to the skin lesion class.

As illustrated in Fig. 2, with the binary mask ($M$) generated from the segmentation task, we are then able to incorporate the clinical knowledge by analyzing the morphological expressions and statistical information from the lesion center area and also the periphery region. {\color{black}Specifically, to locate the center and periphery regions of the skin lesions, we utilize the mathematical morphology analysis on the binary mask $M$:

\begin{equation}
R_{l} = M\ominus E,
\end{equation}
\begin{equation}
R_{m} = M\oplus E-M\ominus E,
\end{equation}

\noindent where $R_{l}$ and $R_{m}$ respectively represent the center region and the periphery region of the skin lesion. $\oplus$ and $\ominus$ are the morphological dilation and erosion operations with the structuring element $E$. Morphological dilation adds pixels to the boundary of the skin lesions, while morphological erosion removes pixels belonging to the skin lesions. The size and shape of structuring element $E$ decide the number of pixels added or removed from the skin lesions.  By combining morphological dilation and erosion into Eqs. (2) and (3), we determine the center and periphery regions of the skin lesions.} The shape delineations obtained using Eqs. (2) and (3) are illustrated as yellow and green color regions, respectively, in Fig. \ref{fig:3}.


For each feature map $f_{k}$ (spatial dimension: $H\times W$, $k\in\{1,2,...,K\}$) of $F$, we propose to squeeze the specific spatial information of the lesion center region and the periphery region separately, which is achieved by extracting statistical information from the feature map ($f_{k}$) as follows:
\begin{equation}
\begin{aligned}
z^{1}_{k} &= \sum_{(x,y)\in R_{l}} f_{k}(x,y)/N_{l},\\
z^{2}_{k} &= \sum_{(x,y)\in R_{m}} f_{k}(x,y)/N_{m},\\
z^{3}_{k} &= \sqrt{\sum_{(x,y)\in R_{l}} (f_{k}(x,y)-z^{1}_{k})^{2}/N_{l}},\\
z^{4}_{k} &= \sqrt{\sum_{(x,y)\in R_{m}} (f_{k}(x,y)-z^{2}_{k})^{2}/N_{m}},
\end{aligned}
\end{equation}

\noindent where $N_{l}$ and $N_{m}$ denote the numbers of pixels in the lesion center region ($R_{l}$) and the periphery region ($R_{m}$), 
respectively. $z^{1}_{k}$ and $z^{2}_{k}$ represent the spatial average information of the lesion center and periphery regions. $z^{3}_{k}$ and $z^{4}_{k}$ measure the standard deviations of feature distribution information of the lesion center and periphery regions.
Hence, for each feature map $f_{k}$ of $F$, we generate a lesion feature descriptor ($z_{k}$) that explicitly characterizes the lesion center and periphery regions, i.e., $z_{k}=[z^{1}_{k},z^{2}_{k},z^{3}_{k},z^{4}_{k}]$. Compared with the feature information obtained by direct global average pooling over feature maps, our final feature representation $\{z_{k}\}_{k=1}^K$ encodes informative statistical information about the morphological expression from both the lesion center area and the periphery region, 
which is much more powerful for melanoma recognition. {\color{black} By feeding $\{z_{k}\}_{k=1}^K$ into a softmax classifier, we obtain the skin lesion classification information $\{g_{r}\}_{r=1}^C$. Here, $\{g_{r}\}_{r=1}^C$ indicates the probability of the lesion belonging to each lesion type. $r$ indexes the skin lesion type (melanoma or non-melanoma in this work), and $C$ is the number of lesion types.}



\subsection{Diagnosis Guided Feature Fusion}

Skin lesion segmentation is a pixel-wise image classification task, i.e., each pixel of a dermoscopic image is assigned with the class label of either lesion or background. Generally, melanoma and non-melanoma lesions have different pathological features. For example, compared with non-melanoma lesions, melanoma cases have more inhomogeneous textures and irregular borders. Existing deep models \cite{yuan2017automatic,tu2019dense,bi2017dermoscopic,al2018skin,li2018dense,nasr2019dense,wang2019bi} often do not consider the lesion type information when learning features for skin lesion segmentation. However, if melanoma and non-melanoma lesions are represented by the feature maps indiscriminately, it will be difficult for the networks to produce effective segmentation for different types of skin lesions. In this paper, we propose a diagnosis guided feature fusion (DGFF) module to incorporate the diagnosis information generated from the melanoma recognition task into the skin lesion segmentation task, which assists our network to produce discriminative feature representations for melanoma and non-melanoma lesions. Specifically, with the guidance of the lesion diagnosis information, the network learns the weights of different feature channels for melanoma and non-melanoma lesions, which thus enables our network to adaptively select the appropriate feature representations for the input image.

To let the diagnosis signals guide the feature learning process of skin lesion segmentation, we first fuse the lesion class information $\{g_{r}\}_{r=1}^C$ with the feature channel-wise information $\{h_{k}\}_{k=1}^K$ of $F$ as:
\begin{equation}
\bm{\mu}= \mathfrak{F}_{e} [\{h_{k}\}_{k=1}^K,\{g_{r}\}_{r=1}^C] ,
\end{equation}

\noindent where $\mathfrak{F}_{e}$ is the concatenation operation. For each feature channel $f_{k}$ of $F$, its feature channel-wise information $h_{k}$ is formulated as:
\begin{equation}
h_{k} = \sum_{(x,y)\in f_{k}} f_{k}(x,y)/N_{k},
\end{equation}

\noindent where $N_{k}$ is the number of pixels in $f_{k}$.
We then normalize $\bm{\mu}$ as follows:
{\color{black}
\begin{equation}
\bm{\tilde{\mu}} = {\textrm{tanh}}(\mathfrak{F}_{t}(\bm{\mu};\theta_{t})),
\end{equation}}
\noindent where $\mathfrak{F}_{t}$ is a linear mapping function to reduce the dimensionality of $\bm{\mu}$ from ($K+C$) to $K$. $\theta_{t}$ are learnable parameters. The obtained $\bm{\tilde{\mu}}$ containing diagnosis signals is used as the channel weight for feature learning. Concretely, we weight each feature channel of $F$ with the guidance of the diagnosis information to generate the feature representation $\bm{\tilde{\mu}}\otimes F$, where $\otimes$ is the channel-wise multiplication.

\begin{figure}[t]
\centering
\includegraphics[width=12 cm,height=4.8cm ]{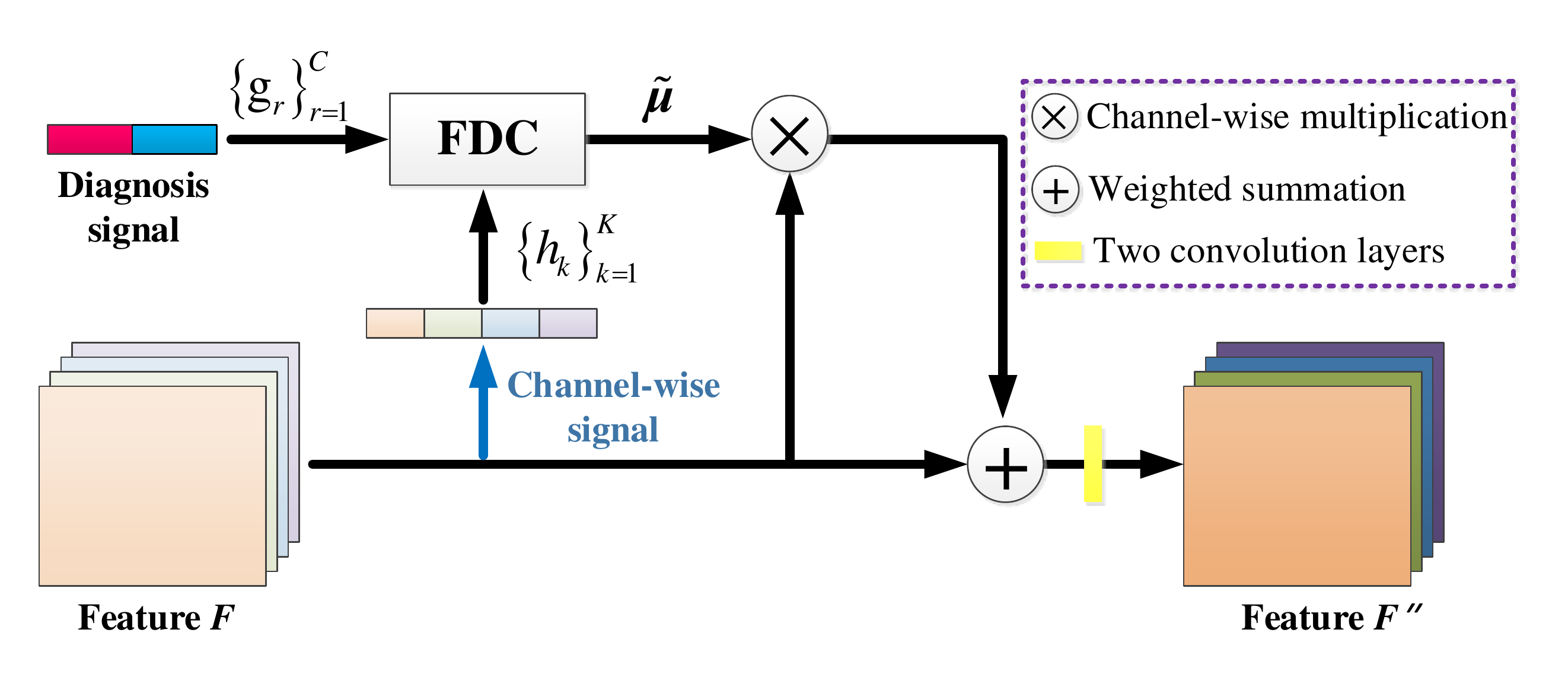}
\vspace*{-0.5cm}
\caption{Illustration of diagnosis guided feature fusion. FDC is the fusion of diagnosis and channel-wise signals.
}
\label{fig:4}
\end{figure}

Therefore, with the guidance of diagnosis information that contains lesion type information, we obtain a set of discriminative feature representations as follows:
\begin{equation}
F^{'} = \mathfrak{F}_{r}(\bm{\tilde{\mu}}\otimes F,F),
\end{equation}

\noindent where the function $\mathfrak{F}_{r}$ learns a weighted summation of $\bm{\tilde{\mu}}\otimes F$ and $F$. After that, we generate the final feature representation $F^{''}$ by feeding $F^{'}$ to two convolutional layers with $3\times3$ kernel size, which further refines the feature fusion information for skin lesion segmentation. $F^{''}$ is then used for pixel-wise skin lesion segmentation. Fig. \ref{fig:4} shows the process of the proposed diagnosis guided feature fusion, where the lesion class information obtained from the melanoma recognition task is embedded into the skin lesion segmentation task. With the guidance of the diagnosis signals from the melanoma recognition, our network is able to adaptively select the discriminative feature representations for handling segmentation on both melanoma images and non-melanoma images.



\subsection{Recursive Mutual Learning}

Skin lesion segmentation can be considered as a pixel-wise image classification process, while melanoma recognition is an image-level classification process. Given pixel-level annotated lesion images and image-level annotated lesion images, the aim of our proposed method is to collaboratively learn skin lesion segmentation and melanoma recognition, in which the two tasks work together to enhance the performance of each other. To promote the mutual learning of skin lesion segmentation and melanoma recognition, we design a recursive mutual learning method to systematically exploit mutual guidance information between these two tasks.

Inspired by the success of recursive network design \cite{yu2018recurrent,zhao2019recursive}, we update the results of skin lesion segmentation and melanoma recognition by incorporating the information generated at the previous stages of the network. Specifically, in our method, the information generated at the previous stage serves as the input for the next stage of our network. This means our proposed architecture iteratively optimizes the two tasks during mutual learning, i.e., the refined lesion segmentation and melanoma recognition are alternatively fed into each other to progressively improve the network's representation capability. For example, let $F^{''}_{n-1}$ be the input feature information at the ($n-1$)\textit{th} stage (as shown in Fig. 2), we enhance the representation ability of $F_{n}$ by designing the recursive counterparts:

\begin{equation}
F^{''}_{n} = \mathfrak{F}_{q} (F^{''}_{n-1}),
\end{equation}

\noindent where $F^{''}_{n}$ represents the output feature information after the recursive counterpart. $\mathfrak{F}_{q}$ is the mutual learning process within the recursive counterpart. Our recursive mutual learning seamlessly integrates skin lesion segmentation and melanoma recognition into a unified framework. It deepens the learning level of the network and benefits the collaborative integration of skin lesion segmentation and melanoma recognition. To effectively train our network with recursive mutual learning, we first train the skin lesion segmentation part until the segmentation performance is accurate and stable. {\color{black} This process ensures a good initialization of the network, which further enhances the recursive mutual learning of the network on skin lesion segmentation and melanoma recognition.} Then we train the skin lesion segmentation and melanoma recognition collaboratively, so that the whole network can converge efficiently for both tasks.


\section{Experimental Evaluation}

\subsection{Materials}
We evaluate our proposed framework on two public benchmark datasets (ISBI 2016~\cite{timmurphy16} and ISBI 2017~\cite{timmurphy17}). They are provided by the International Skin Imaging Collaboration (ISIC) for the International Symposium on Biomedical Imaging challenges named ``Skin Lesion Analysis toward Melanoma Detection''. ISBI 2016 dataset includes a training set with 900 dermoscopic images (727 non-melanoma cases and 173 melanoma cases), and a testing set with 379 dermoscopic images (304 non-melanoma cases and 75 melanoma cases). ISBI 2017 dataset is an extension of the ISBI 2016 dataset, which contains three different sets for training, validation and test. Training set comprises 2000 dermoscopic images, i.e., 374 melanoma cases, 254 seborrheic keratosis cases, and 1372 benign nevi cases. Validation set contains 150 annotated dermoscopic images, i.e., 30 melanoma cases, 42 seborrheic keratosis cases, and 78 benign nevi cases. Test set consists of 600 dermoscopic images, i.e., 117 melanoma cases, 90 seborrheic keratosis cases, and 393 benign nevi cases.

\subsection{Evaluation Criteria}
\textbf{Skin lesion segmentation:} We adopt four different metrics for skin lesion segmentation evaluation, including jaccard index (JA), dice coefficient (DI), segmentation accuracy ($\rm{AC}_{s}$) and geometric mean (GE). JA and DI are two metrics that estimate the similarity between the ground truths and the segmented skin lesions: JA = $TP/(TP+FN+FP)$ and DI = $2TP/(2TP+FN+FP)$, where $TP$ and $TN$ are the number of pixels correctly segmented skin lesion and background pixels; $FN$ and $FP$ are the number of pixels incorrectly segmented background and lesion pixels. $\rm{AC}_{s}$ is the ratio of the number of correctly segmented lesion and background pixels to the total number of pixels. GE is the average value of sensitivity and specificity of the segmentation performance. Jaccard index (JA) was taken as the most important criterion for skin lesion segmentation comparison in ISBI 2016 and 2017 challenges.

\textbf{Melanoma recognition:} To evaluate the melanoma recognition performance, we employ four measurements including the melanoma recognition accuracy ($\rm{AC}_{r}$), sensitivity (SE), specificity (SP) and the area under the receiver operating characteristic curve (AUC). $\rm{AC}_{r}$ is the ratio of the number of correctly recognized melanoma and non-melanoma cases to the total number of lesion cases. SE represents the ratio of the number of correctly recognized melanoma to the total number of melanoma cases. SP denotes the ratio of the number of correctly recognized non-melanoma to the total number of non-melanoma cases. The receiver operating characteristic (ROC) curve is plotted with true positive fraction (sensitivity) versus false positive fraction (1-specificity) by varying the threshold on the probability map. AUC measures the area under the ROC curve. ISBI 2016 and 2017 challenges took AUC as the most important criterion for melanoma recognition comparison.

\subsection{Implementation Details}
We use the ResNet101 model~\cite{he2016deep} (pre-trained on ImageNet~\cite{russakovsky2015imagenet}) as our baseline network. To optimize the skin lesion segmentation and melanoma recognition tasks, we use both the pixel-level annotated data and the image-level annotated data for model training. Two cross-entropy losses $L_{seg}$ and $L_{cls}$ are adopted to minimize the distance between the output of each task and the corresponding ground truth. {\color{black} The total loss is a linear combination of the skin lesion segmentation and melanoma recognition: \emph{Loss}=$L_{seg}$+$\beta$$L_{cls}$. Here, $\beta$ is the parameter to control this multi-task loss, which is empirically set to be 0.3 in this work.} The proposed network is trained end-to-end using standard stochastic gradient descent with batch size 8. For skin lesion segmentation, a deconvolutional layer is exploited to increase the size of score maps to that of the input images~\cite{ding2018context}. {\color{black} We denote the feature generation process (before global average pooling ) of the baseline network as five blocks, i.e., Block 1, Block2, $\dots$, Block 5, where each Block produces the feature maps with different resolutions. For melanoma recognition task, the features are learned from the cascade stages of the network. Since features from shallow layers have more appearance information that help the skin lesion segmentation, we apply the skip layers to extract multiscale information from the shallow layers to increase the skin lesion segmentation performance as~\cite{long2015fully}. Specifically, we exploit the FCN-8s~\cite{long2015fully} architecture where the outputs from three Blocks (Block 3, Block 4, Block 5) are up-sampled and fused with features from cascade stages to generate the final feature maps for skin lesion segmentation.} {\color{black} For morphological analysis in Eqs. (2) and (3), we create a disk-shaped structuring element $E$, whose radius is set to be 1/16 of the height of binary mask $M$.} Following~\cite{ding2020semantic,ding2019boundary}, we first set the initial learning rate to $10^{-3}$ and then update it by the ``poly" learning rate policy. For batch processing, each image is resized to have maximum extent of 512 pixels.



As ISBI 2016 dataset only provides training and test sets for skin lesion segmentation and melanoma recognition, we randomly pick up 800 images from training set for training and the rest 100 images from training set for validation. For ISBI 2017 dataset, training and validation are performed on its provided training and validation sets. We set the numbers of iterations for ISBI 2016 and ISBI 2017 datasets to 30k and 60k respectively. Data augmentation strategies, like random flipping and cropping, random scaling in a range of [0.8, 1.2], are applied to further enlarge the training dataset.


\subsection{Ablation Studies}

\subsubsection{Evaluation of Each Component in Our Framework}

To investigate the effectiveness of our approach on skin lesion segmentation and melanoma recognition, we conduct comprehensive studies on the challenging ISBI 2016 and ISBI 2017 datasets. Tables \ref{tab:1} and \ref{tab:2} show the performance of our proposed method on skin lesion segmentation and melanoma recognition respectively. Compared with the baseline model for skin lesion segmentation, the model integrating the lesion class information from the melanoma recognition task enhances the skin lesion segmentation performance, i.e., improving the JA by 2.9\%-4.4\% on ISBI 2016 and ISBI 2017 datasets. In contrast to the baseline model for melanoma recognition, the model embedding the lesion structure information from the skin lesion segmentation task achieves a significant performance increase of melanoma recognition, i.e., resulting in the 1.7\%-5.2\% of AUC improvement on ISBI 2016 and ISBI 2017 datasets. By learning the mutual information from both skin lesion segmentation task and melanoma recognition task in a recursive manner, the proposed method in 3 recursive stages further boosts the skin lesion segmentation and melanoma recognition performance on ISBI 2016 and ISBI 2017 datasets consistently. For example, we achieve 1.7\%-2.9\% of JA enhancement for skin lesion segmentation, and 1.7\%-2.4\% of AUC enhancement for melanoma recognition on ISBI 2016 and ISBI 2017 datasets respectively.

\begin{table}[t]
\centering
\setlength{\abovecaptionskip}{0pt}
\caption{Performance of skin lesion segmentation on ISBI 2016 and ISBI 2017 datasets (\%)}
\renewcommand\arraystretch{0.65}
\setlength{\tabcolsep}{6pt}
\label{tab:1}
\begin{tabular}{llllll}
\toprule
\toprule
\multicolumn{1}{l}{Dataset} &\multicolumn{1}{l}{Method} &\multicolumn{1}{c}{JA} &\multicolumn{1}{c}{$\rm{AC}_{s}$} &\multicolumn{1}{c}{DI} &\multicolumn{1}{c}{GM}\\
\midrule
&Baseline  &84.5 &95.6 &90.8 &93.8\\
ISBI 2016&Baseline+DGFF &87.4 &96.4 &92.8 &95.1\\
&\textbf{Proposed method} &\textbf{89.1} &\textbf{96.9} &\textbf{93.9} &\textbf{95.5}\\
\midrule
&Baseline &75.1 &92.7 &83.5 &86.4  \\
ISBI 2017&Baseline+DGFF &79.5 &94.2 &87.4 &88.4\\
&\textbf{Proposed method} &\textbf{82.4} &\textbf{95.2} &\textbf{89.4} &\textbf{91.0}\\
\bottomrule
\bottomrule
\end{tabular}
\end{table}

\begin{table}[t]
\centering
\setlength{\abovecaptionskip}{0pt}
\caption{Performance of melanoma recognition on ISBI 2016 and ISBI 2017 datasets (\%)}
\renewcommand\arraystretch{0.65}
\setlength{\tabcolsep}{6pt}
\label{tab:2}
\begin{tabular}{llllll}
\toprule
\toprule
\multicolumn{1}{l}{Dataset} &\multicolumn{1}{l}{Method} &\multicolumn{1}{c}{AUC} &\multicolumn{1}{c}{$\rm{AC}_{r}$} &\multicolumn{1}{c}{SE}&\multicolumn{1}{c}{SP}\\
\midrule
&Baseline  &83.6 &83.6 &\textbf{57.3} &83.6\\
ISBI 2016&Baseline+LPSE &85.3 &85.5 &41.3 &\textbf{96.4}\\
&\textbf{Proposed method} &\textbf{87.7} &\textbf{87.3} &52.0 &96.1 \\
\midrule
&Baseline &82.2 &83.8 &42.7 &93.8  \\
ISBI 2017&Baseline+LPSE &87.4 &86.9 &58.2 &93.5\\
&\textbf{Proposed method} &\textbf{89.1} &\textbf{88.2} &\textbf{63.3} &\textbf{94.2}\\
\bottomrule
\bottomrule
\end{tabular}
\end{table}

\begin{figure}[t]
\hspace*{-0.2cm}
\includegraphics[width=13 cm,height=4cm ]{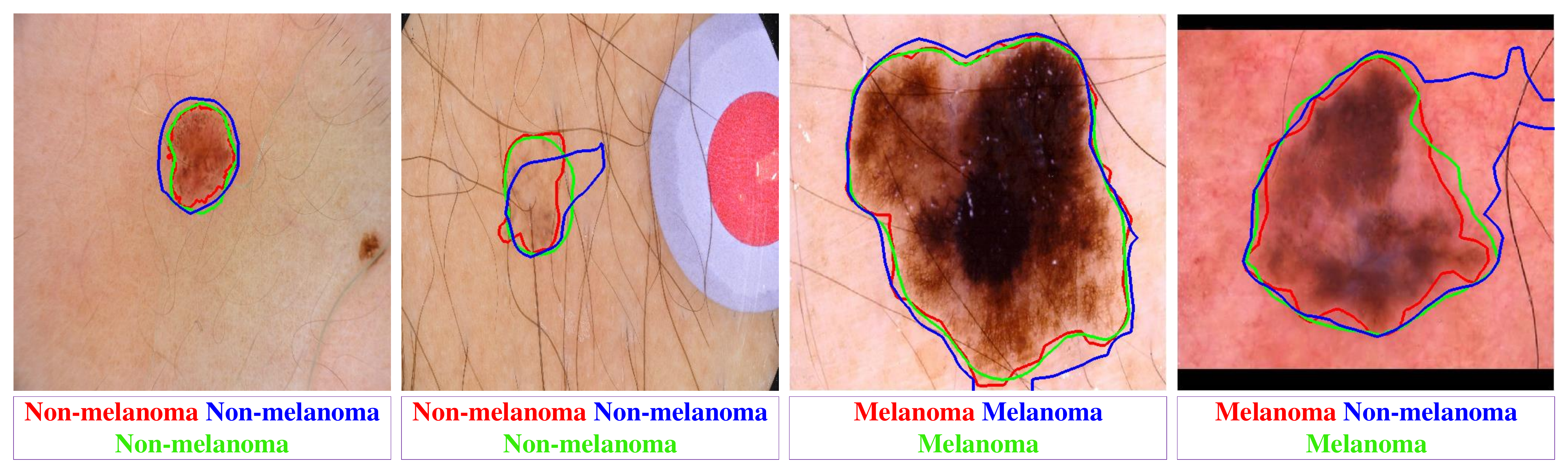}
\vspace*{-0.7cm}
\caption{Qualitative evaluation of skin lesion segmentation and melanoma recognition on ISBI 2016 dataset. First row: skin lesion segmentation results of ground truth (red contour), baseline segmented model (blue contour) and our proposed framework (green contour) respectively. Second row: melanoma recognition results of ground truth (red font), baseline recognition model (blue font) and our proposed framework (green font) respectively.}
\label{fig:516}
\end{figure}

\begin{figure}[t]
\hspace*{-0.2cm}
\includegraphics[width=13 cm,height=4cm ]{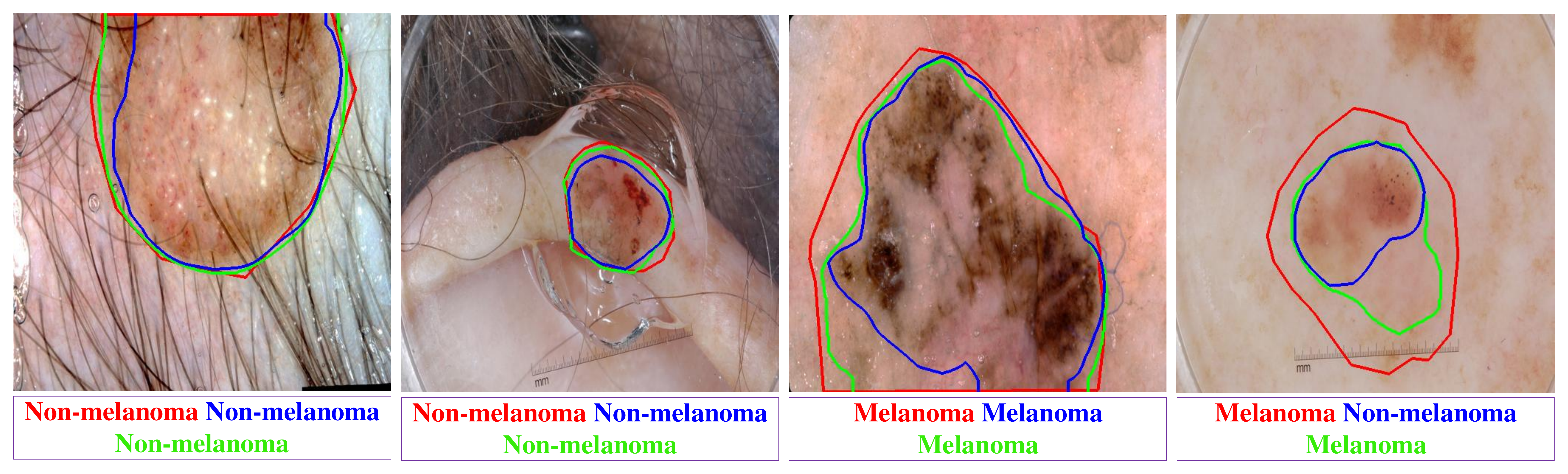}
\vspace*{-0.7cm}
\caption{Qualitative evaluation of skin lesion segmentation and melanoma recognition on ISBI 2017 dataset. First row: skin lesion segmentation results of ground truth (red contour), baseline segmented model (blue contour) and our proposed framework (green contour) respectively. Second row: melanoma recognition results of ground truth (red font), baseline recognition model (blue font) and our proposed framework (green font) respectively.}
\label{fig:517}
\end{figure}

We also qualitatively show the performance of our proposed framework on skin lesion segmentation and melanoma recognition in Figs. \ref{fig:516} and \ref{fig:517}. We display some challenging skin lesion cases in ISBI 2016 and ISBI 2017 datasets, like the lesion with different appearance features (e.g., color, shape, texture), the lesion with low contrast, and lesion images with some noises. Compared with the baseline models that segment skin lesion and recognize melanoma separately, our proposed framework can simultaneously achieve skin lesion segmentation and melanoma recognition. For skin lesion segmentation, the proposed framework compared with the baseline segmentation, produces better visualization of skin lesion segmentation including more accurate delineation of skin lesions with different shape, color and texture characteristic. For melanoma recognition, the proposed framework compared with the baseline recognition, achieves better melanoma detection performance. For example, the last columns in Figs. \ref{fig:516} and \ref{fig:517} show the challenging melanoma cases. The baseline model for recognition is not able to classify those types of skin lesions as melanoma, while our proposed framework correctly classifies them to be melanoma.


\subsubsection{Evaluation of Different Numbers of Cascade Stages}

To evaluate the performance of different numbers of cascade stages on recursive mutual learning process, we design the experiments of skin lesion segmentation and melanoma recognition at multiple learning stages. Since ISBI 2017 dataset has more challenging melanoma and non-melanoma cases, we take ISBI 2017 dataset as an illustration for the evaluation of different numbers of cascade stages. {\color{black} Table \ref{tab:3} shows the results of different numbers of cascade stages for skin lesion segmentation and melanoma recognition. From Table \ref{tab:3}, we find the proposed framework improves the performance of skin lesion segmentation and melanoma recognition progressively with the increase of the numbers of cascade stages. This further exemplifies the consistent effectiveness of the proposed method on skin lesion diagnosis. When the cascade stage number is larger than 3, the performance of skin lesion segmentation and melanoma recognition is slightly enhanced. Considering the computation time and GPU memory, we take the performance of the proposed network with three cascade stages as the example for the following analysis and comparison. } 

\begin{table}[t]
\centering
\hspace*{0.6cm}
\setlength{\abovecaptionskip}{0pt}
\caption{Performance of different cascade numbers of the proposed network on ISBI 2017 dataset (\%)}
\renewcommand\arraystretch{0.65}
\setlength{\tabcolsep}{6pt}
\label{tab:3}
\begin{tabular}{cllllllll}
\toprule
\toprule
\multicolumn{1}{c}{\multirow{2}{*}{Number}} &\multicolumn{4}{c}{Skin lesion segmentation} &\multicolumn{4}{c}{Melanoma recognition} \\
\cline{2-5} \cline{6-9}
 &\multicolumn{1}{c}{JA} &\multicolumn{1}{c}{$\rm{AC}_{s}$} &\multicolumn{1}{c}{DI} &\multicolumn{1}{c}{GM} &\multicolumn{1}{c}{AUC} &\multicolumn{1}{c}{$\rm{AC}_{r}$} &\multicolumn{1}{c}{SE} &\multicolumn{1}{c}{SP} \\
\midrule
1 &79.5 &94.2 &87.4 &88.4 &87.4 &86.9 &58.2 &93.5\\
2 &81.1 &94.8 &88.4 &89.8 &88.0 &87.0 &51.3 &95.6\\
3 &82.4 &95.2 &89.4 &91.0 &89.1 &88.2 &63.3 &94.2\\
4 &82.7 &95.2 &89.6 &91.9 &89.0 &88.0 &62.4 &94.2\\
5 &82.9 &95.3 &89.8 &92.1 &89.2 &87.3 &59.0 &94.1\\
\bottomrule
\bottomrule
\end{tabular}
\end{table}

\begin{figure}[htbp]
\hspace*{-0.2cm}
\includegraphics[width=13 cm,height=4cm ]{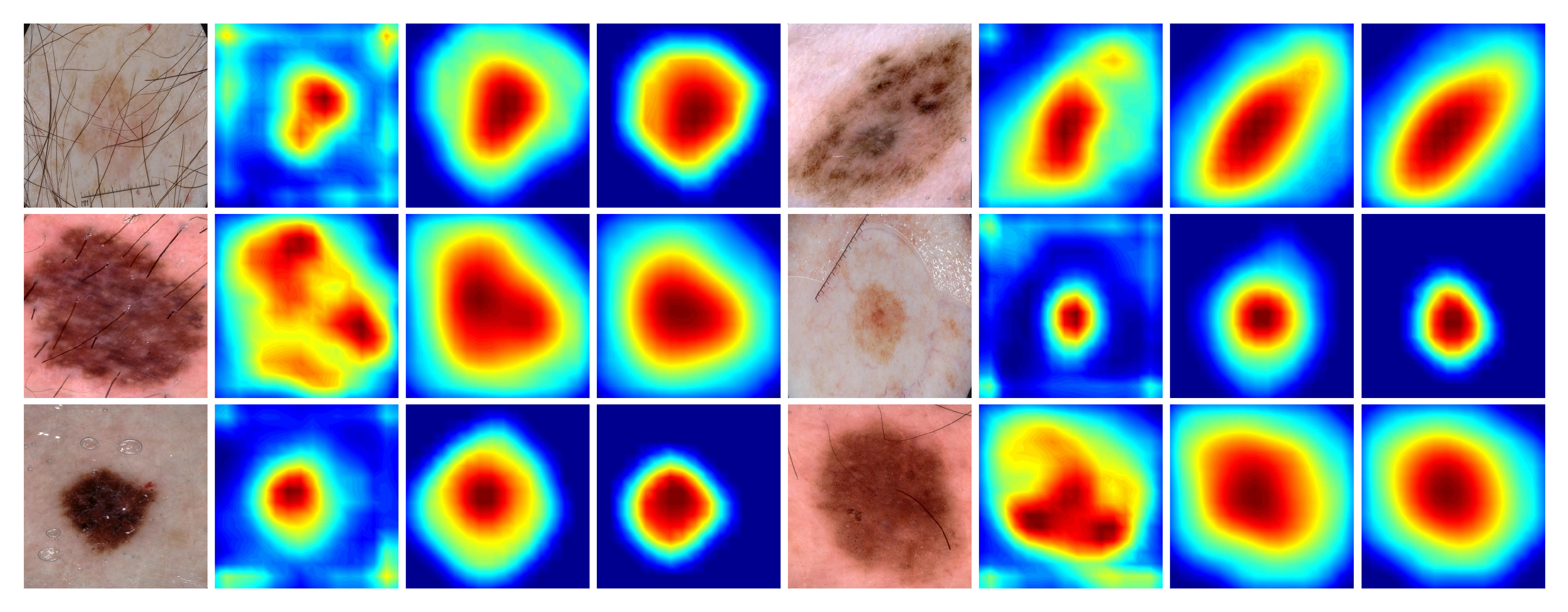}
\vspace*{-0.7cm}
\caption{{\color{black}Visualization examples of saliency images. First and fifth columns: original skin lesion images. Second to fourth columns: the saliency maps of features of original images (first column) at the 1st, 2nd and 3rd cascade stages, respectively. Sixth to eighth columns: the saliency maps of features of original images (fifth column) at the 1st, 2nd and 3rd cascade stages, respectively.}}
\label{fig:add}
\end{figure}

{\color{black} We also visualize the saliency maps of features from three different cascade stages by GradCAM~\cite{selvaraju2017grad}, as shown in Fig. \ref{fig:add}. It can be observed that our proposed network can learn to focus on the discriminative parts of skin lesions. For some challenging dermoscopic images with noises (e.g., hairs and rulers) and low contrast, the proposed network also shows the promising localization ability for the skin lesions. Furthermore, with the increase of the cascade stage number, our network concentrates more on the specifically informative parts of skin lesions, which displays the effectiveness of the recursive learning on skin lesion analysis. In addition, the detailed visualization in Fig. \ref{fig:add} also makes the learning process of our proposed network more explainable.}
\subsection{Comparison with Other Methods}

\begin{table}[t]
\centering
\hspace*{0.6cm}
\setlength{\abovecaptionskip}{0pt}
\caption{Performance comparison of skin lesion segmentation on ISBI 2016 dataset (\%)}
\renewcommand\arraystretch{0.65}
\setlength{\tabcolsep}{6pt}
\label{tab:comseg16}
\begin{tabular}{llcccccc}
\toprule
\toprule
\multicolumn{1}{l}{\multirow{2}{*}{Method}} &\multicolumn{2}{c}{Melanoma} &\multicolumn{2}{c}{Non-melanoma} &\multicolumn{2}{c}{Overall}\\
\cline{2-7}
 &\multicolumn{1}{c}{JA} &\multicolumn{1}{c}{$\rm{AC}_{s}$} &\multicolumn{1}{c}{JA} &\multicolumn{1}{c}{$\rm{AC}_{s}$} &\multicolumn{1}{c}{JA} &\multicolumn{1}{c}{$\rm{AC}_{s}$}\\
\midrule
Team-EXB~\cite{timmurphy16} &82.9 &93.2 &84.6 &95.8  &84.3 &95.3 \\
Team-CUMED~\cite{yu2017automated} &82.9 &93.2 &83.0 &95.3 &82.9 &94.9  \\
Team-Rahman~\cite{timmurphy16} &82.7 &93.2 &82.0 &95.7 &82.2 &95.2 \\
Team-SFU~\cite{timmurphy16} &81.9 &92.2 &80.9 &94.9 &81.1 &94.4 \\
Team-TMU~\cite{timmurphy16} &82.3 &93.4 &80.7 &94.9 &81.0 &94.6 \\
\midrule
Yuan \emph {et al}.~\cite{yuan2017automatic}&N.A &N.A &N.A &N.A &84.7 &95.5 \\
Yuan \emph {et al}.~\cite{yuan2017improving}&N.A &N.A &N.A &N.A &84.9 &95.7\\
Li \emph {et al}.~\cite{li2018dense} &N.A &N.A &N.A &N.A &87.0 &95.9 \\
Bi \emph {et al}.~\cite{bi2017dermoscopic} &85.8 &94.7 &84.3 &95.7 &84.6 &95.5\\
Bi \emph {et al}.~\cite{bi2019step} &85.6 &94.3 &85.6 &96.2 &85.9 &95.8 \\
Wang \emph {et al}.~\cite{wang2019bi} &88.4 &95.9 &88.1 &97.0 &88.1 &96.7\\
\textbf{Proposed method} &\textbf{89.3} &\textbf{96.0} &\textbf{89.0} &\textbf{97.2} &\textbf{89.1} &\textbf{96.9}\\
\bottomrule
\bottomrule
\end{tabular}
\end{table}

\begin{table}[h]
\centering
\hspace*{0.6cm}
\setlength{\abovecaptionskip}{0pt}
\caption{Performance comparison of skin lesion segmentation on ISBI 2017 dataset (\%)}
\renewcommand\arraystretch{0.65}
\setlength{\tabcolsep}{6pt}
\label{tab:comseg17}
\begin{tabular}{llcccccc}
\toprule
\toprule
\multicolumn{1}{l}{\multirow{2}{*}{Method}} &\multicolumn{2}{c}{Melanoma} &\multicolumn{2}{c}{Non-Melanoma} &\multicolumn{2}{c}{Overall}\\
\cline{2-7}
 &\multicolumn{1}{c}{JA} &\multicolumn{1}{c}{$\rm{AC}_{s}$} &\multicolumn{1}{c}{JA} &\multicolumn{1}{c}{$\rm{AC}_{s}$} &\multicolumn{1}{c}{JA} &\multicolumn{1}{c}{$\rm{AC}_{s}$}\\
\midrule
Team-Yuan~\cite{yuan2017improving} &71.2 &90.0 &77.8 &94.2 &76.5 &93.4 \\
Team-Berseth~\cite{timmurphy17} &68.8 &89.0 &78.0 &94.2 &76.2 &93.2\\
Team-popleyi~\cite{timmurphy17} &69.3 &89.6 &77.6 &94.3 &76.0 &93.4 \\
Team-Ahn~\cite{timmurphy17} &69.1 &89.6 &77.5 &94.3 &75.8 &93.4\\
Team-RECOD~\cite{timmurphy17}  &68.8 &89.4 &77.0 &94.0 &75.4 &93.1 \\
\midrule
Li \emph {et al}.~\cite{li2018dense} &N.A &N.A &N.A &N.A &76.5 &93.9 \\
Tu \emph {et al}.~\cite{tu2019dense} &N.A &N.A &N.A &N.A &76.8 &94.5\\
Bi \emph {et al}.~\cite{bi2019step} &72.2 &90.1 &79.1 &95.1 &77.7 &94.1 \\
Wang \emph {et al}.~\cite{wang2019bi} &77.3 &92.0 &82.5 &95.3 &81.5 &94.7\\
Xie \emph {et al}.~\cite{xie2020mutual} &N.A &N.A &N.A &N.A &80.4 &94.7\\
\textbf{Proposed method} &\textbf{77.4} &\textbf{92.2} &\textbf{83.7} &\textbf{95.9} &\textbf{82.4} &\textbf{95.2}\\
\bottomrule
\bottomrule
\end{tabular}
\end{table}

We compare our proposed framework with other state-of-art methods on skin lesion segmentation and melanoma recognition. Tables \ref{tab:comseg16} and \ref{tab:comseg17} shows the performance comparison of skin lesion segmentation, including the top 5 segmentation results on ISBI 2016 and 2017 skin lesion segmentation challenge. Compared with the other methods on skin lesion segmentation, the proposed method achieves the highest skin lesion segmentation performance, i.e., producing the best JA for all skin lesion cases in ISBI 2016 and 2017 datasets. For specific lesion cases like melanoma and non-melanoma, the proposed method also generates the better skin lesion segmentation results consistently for different types of skin lesions. 

In Table \ref{tab:comreg16}, we compare the performance of the proposed framework with different deep learning methods on melanoma recognition, including the top 5 melanoma recognition results on ISBI 2016 skin lesion classification challenge. In contrast to other melanoma recognition methods, the proposed knowledge-aware deep framework provides a noticeable improvement of the melanoma recognition, i.e., increasing the AUC by 2.5\% from the best reported result~\cite{yu2018melanoma}. Table \ref{tab:comreg17} shows the comparison of different deep learning methods on melanoma recognition for ISBI 2017 dataset. Note that some results reported in the literature are produced by training the networks using many additional external skin lesion images to enhance the performance, like the top five techniques in~\cite{timmurphy17}, and the methods in~\cite{zhang2019attention,xie2020mutual}. For example, $\rm{AUC}$ in \cite{zhang2019attention} is increased from 85.9\% as shown in Table \ref{tab:comreg17} to 87.5\% by using the additional training data. These results are not comparable to ours, since they use additional training samples not specified to train their networks. To make our results reproducible by others and comparable with others, our proposed framework is trained only by the standard training samples that are from the training set (2000 images) of ISBI 2017 challenge without any other external training data. From Table \ref{tab:comreg17}, we can observe that the proposed method achieves the best AUC, i.e., 2.0\% higher than the best reported result~\cite{liang2019multi}. In addition, the high melanoma recognition with less training data achieved by the proposed framework also provides a potential way to extend the proposed framework to other medical image analysis application with small training sample size.

{\color{black}In addition, the method in~\cite{wang2019bi} is our previous work for skin lesion segmentation. Main differences between the method in~\cite{wang2019bi} and our new proposed network are concluded as follows: first, our new proposed network presents a novel collaborative learning strategy for skin lesion segmentation and melanoma recognition, where the benefits from both tasks are transferred with the mutual guidance to boost the performance of the skin lesion analysis. However, the previous work in~\cite{wang2019bi} didn't use this collaborative learning process for skin lesion segmentation. Second, our new proposed method achieves an explainable diagnosis of the skin lesions by explicitly utilizing the clinical knowledge within deep feature learning process, while the previous work in~\cite{wang2019bi} only learns the lesion feature representation implicitly through the deep framework. Third, we design a recursive mutual learning to enhance the joint learning ability for skin lesion segmentation and melanoma recognition, while the previous work in~\cite{wang2019bi} applies a straightforward learning manner to segment the skin lesion. Compared with the method in~\cite{wang2019bi} for skin lesion segmentation, our proposed network improves JA by 0.9\%-1\% on two publicly available datasets consistently (as shown in Tables \ref{tab:comseg16}-\ref{tab:comseg17}), which exemplifies the effectiveness of incorporating lesion diagnosis knowledge into skin lesion segmentation. Furthermore, we also test the work in~\cite{wang2019bi} for melanoma recognition by utilizing the features from the top convolution layers of the network. Compared with the work in~\cite{wang2019bi} for melanoma recognition, our proposed network enhances AUC by 3.5\%-3.9\% (as shown in Tables \ref{tab:comreg16}-\ref{tab:comreg17}), which shows the benefit of transferring the skin lesion segmentation information to the melanoma recognition.}

\begin{table}[t]
\centering
\hspace*{0.6cm}
\setlength{\abovecaptionskip}{0pt}
\caption{Performance comparison of melanoma recognition on ISBI 2016 dataset (\%)}
\renewcommand\arraystretch{0.65}
\setlength{\tabcolsep}{13pt}
\label{tab:comreg16}
\begin{tabular}{lcccc}
\toprule
\toprule
\multicolumn{1}{l}{Method} &\multicolumn{1}{c}{AUC} &\multicolumn{1}{c}{$\rm{AC}_{r}$} &\multicolumn{1}{c}{SE} &\multicolumn{1}{c}{SP}\\
\midrule
Team-BFTB~\cite{timmurphy16}&82.6 &83.4 &32.0 &96.1 \\
Team-GTDL~\cite{timmurphy16} &80.2 &81.3 &57.3 &87.2\\
Team-GTDL2~\cite{timmurphy16} &80.0 &68.1 &78.7 &65.5\\
Team-GTDL1~\cite{timmurphy16} &81.3 &81.5 &46.7 &90.1\\
Team-CUMED~\cite{timmurphy16} &80.4 &85.5 &50.7 &94.1\\
\midrule
Al \emph {et al}.~\cite{al2020multiple} &76.6 &81.2 &\textbf{81.8} &71.4\\
Yu \emph {et al}.~\cite{yu2018melanoma} &85.2 &86.8  &N.A. &N.A.\\
Yu \emph {et al}.~~\cite{yu2017automated} &80.4 &85.5 &50.7 &94.1 \\
Sultana \emph{et al}.~\cite{sultana2018deep}&83.5 &86.1 &56.0 &92.4 \\
Wang \emph {et al}.~\cite{wang2019bi} &84.2 &84.7 &46.7 &94.1\\
\textbf{Proposed method} &\textbf{87.7} &\textbf{87.3} &52.0 &\textbf{96.1} \\
\bottomrule
\bottomrule
\end{tabular}
\end{table}

\begin{table}[t]
\centering
\hspace*{0.8cm}
\setlength{\abovecaptionskip}{0pt}
\caption{Performance comparison of melanoma recognition without external training data on ISBI 2017 dataset (\%)}
\renewcommand\arraystretch{0.65}
\setlength{\tabcolsep}{13pt}
\label{tab:comreg17}
\begin{tabular}{lcccc}
\toprule
\toprule
\multicolumn{1}{l}{Method} &\multicolumn{1}{c}{AUC} &\multicolumn{1}{c}{$\rm{AC}_{r}$} &\multicolumn{1}{c}{SE} &\multicolumn{1}{c}{SP}\\
\midrule
Zhang \emph {et al}.~\cite{zhang2019attention}  &85.9 &83.7 &59.0 &89.6\\
Yang \emph {et al}.~\cite{yang2018classification}  &84.2 &83.0 &60.7 &88.4\\
Sultana \emph{et al}.~\cite{sultana2018deep} &78.9 &83.2 &52.9 &90.5 \\
Al \emph {et al}.~\cite{al2020multiple} &81.6 &N.A. &75.3 &80.6\\
Harangi~\cite{harangi2018skin}  &85.1 &85.2 &40.2 &71.9\\
Liang \emph{et al}.~\cite{liang2019multi} &87.1 &86.8 &57.0 &\textbf{98.0}\\
Wang \emph {et al}.~\cite{wang2019bi} &85.2 &85.8 &53.9 &93.5\\
\textbf{Proposed method} &\textbf{89.1} &\textbf{88.2} &\textbf{63.3} &94.2 \\
\bottomrule
\bottomrule
\end{tabular}
\vskip1.5mm
\textit{\footnotesize For fair comparison, only results from the same training data are recorded.}
\end{table}

{\color{black}\subsection{Evaluation on ISIC archive dataset}

ISIC archive dataset\footnote{\url{https://www.isic-archive.com/}} supports more skin lesion samples with melanoma diagnosis and lesion segmentation information. We use 13779 skin lesion images to evaluate the performance of our proposed network on skin lesion segmentation and melanoma recognition. Here, we take 80\% of those images as training set, 10\% and 10\% of them as validation and test sets. Table \ref{tab:bigdata} shows the performance of our proposed network on ISIC archive dataset. Compared with the performance of the baseline model, our proposed network improves JA by 6.1\% for skin lesion segmentation, and AUC by 2.8\% for melanoma recognition, which justifies the validity and robustness of the proposed network on skin lesion analysis.

\begin{table}[t]
\centering
\hspace*{0.6cm}
\setlength{\abovecaptionskip}{0pt}
\caption{Skin lesion segmentation and melanoma recognition on ISIC archive dataset (\%)}
\renewcommand\arraystretch{0.65}
\setlength{\tabcolsep}{6pt}
\label{tab:bigdata}
\begin{tabular}{llllll}
\toprule
\toprule

\multicolumn{1}{l}{Task} &\multicolumn{1}{l}{Method} &\multicolumn{1}{c}{JA} &\multicolumn{1}{c}{$\rm{AC}_{s}$} &\multicolumn{1}{c}{DI} &\multicolumn{1}{c}{GM}\\
\midrule
Skin lesion &Baseline &73.4 &96.1 &84.7 &91.8 \\
segmentation &\textbf{Proposed method}  &\textbf{79.5} &\textbf{97.0} &\textbf{86.0} &\textbf{92.4}\\
\midrule
\multicolumn{1}{l}{Task} &\multicolumn{1}{l}{Method} &\multicolumn{1}{c}{AUC} &\multicolumn{1}{c}{$\rm{AC}_{r}$} &\multicolumn{1}{c}{SE} &\multicolumn{1}{c}{SP}\\
\midrule
Melanoma &Baseline &92.7 &92.9 &40.0 &\textbf{97.8}\\
recognition &\textbf{Proposed method} &\textbf{95.5} &\textbf{93.4}  &\textbf{55.4} &97.0\\
\bottomrule
\bottomrule
\end{tabular}
\end{table}}

\subsection{Result Summary}
Skin lesion segmentation and melanoma recognition are two most critical functions for the automatic diagnosis of melanoma. We propose a knowledge-aware deep framework that achieves state-of-the-art performance for skin lesion segmentation and melanoma recognition on two publicly available skin lesion datasets. Experimental results in Tables \ref{tab:1} and \ref{tab:2}, Figs. \ref{fig:516} and \ref{fig:517}, show the effectiveness of the proposed method on skin lesion segmentation and melanoma recognition respectively. We attribute the melanoma recognition performance enhancement to the network embedded with morphological analysis that incorporates the skin lesion structure information from skin lesion segmentation into the melanoma recognition. Thus, the network have capability of learning informative feature containing the useful clinical statistical information of both lesion and its border regions to identify the challenging melanoma. In addition, the improvement of the skin lesion segmentation comes from the integration of the lesion class information from melanoma recognition into skin lesion segmentation, which assists the network to select the discriminative lesion features for different types of lesion cases (i.e., melanoma and non-melanoma) for skin lesion segmentation. Tables \ref{tab:comseg16}-\ref{tab:comreg17} show the comparison of our proposed framework with other methods on skin lesion segmentation and melanoma recognition. Those results clearly show the efficiency of our knowledge-aware deep framework that exploits mutual benefits from both skin lesion segmentation and melanoma recognition to jointly improve each task performance.

\begin{figure}[t]
\centering
\includegraphics[width=9 cm,height=5cm ]{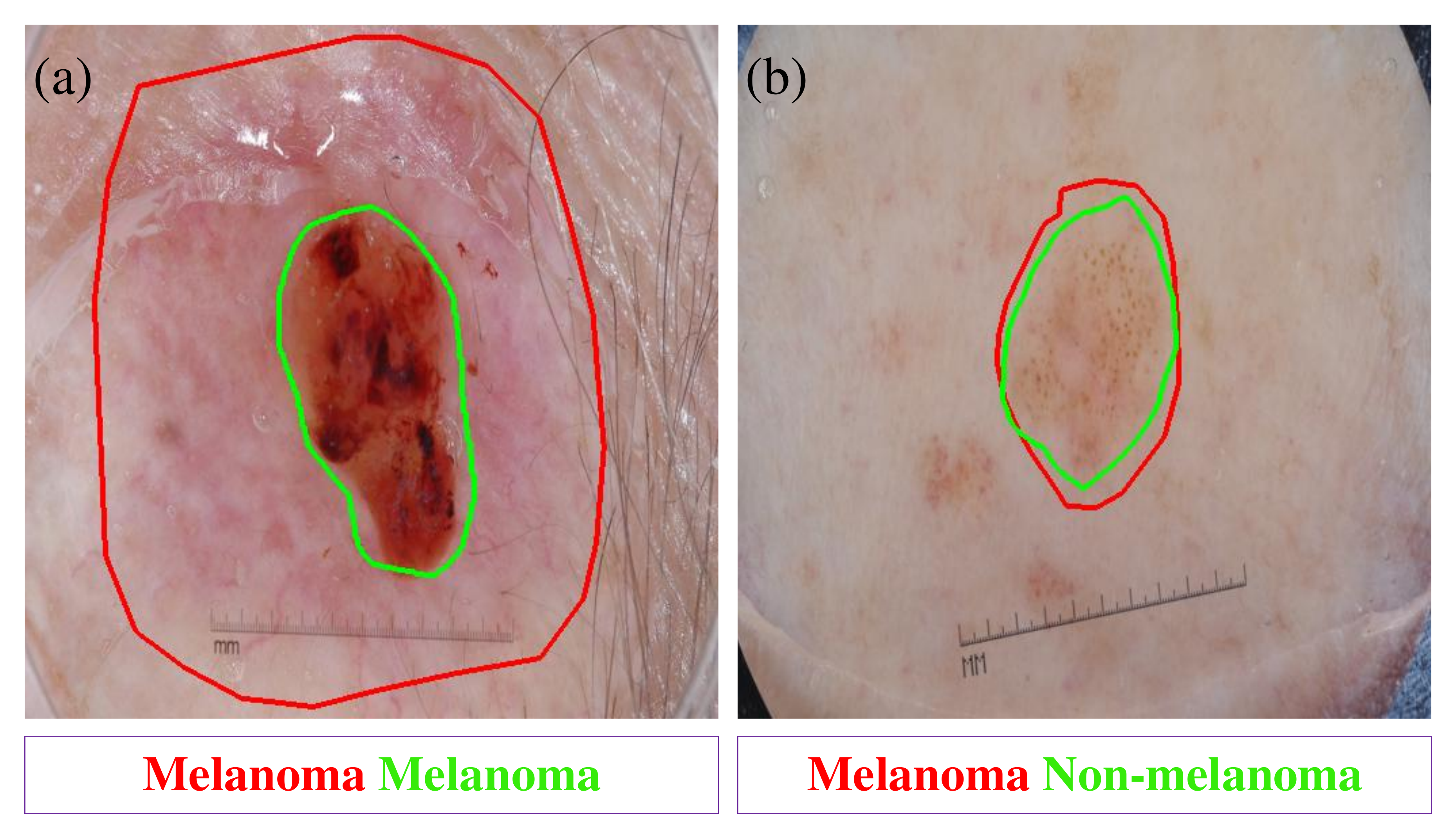}
\caption{Examples of some challenging cases for skin lesion segmentation and melanoma recognition that need further investigation. (a) a challenging case for skin lesion segmentation. (b) a challenging case for melanoma recognition. {\color{black} Red and green contours denote the skin lesion segmentation results of ground truths and the proposed method, while red and green fonts represent the melanoma recognition results of ground truths and the proposed method.}}
\label{fig:7}
\end{figure}


Though the proposed framework have achieved promising performance for skin lesion segmentation and melanoma recognition, there are still some dermoscopic cases that need to be further investigated for both tasks. Fig. \ref{fig:7} (a) shows one failure case for skin lesion segmentation, where the skin lesion has an area of redness on the surrounding of lesion center. Fig. \ref{fig:7} (b) shows the failure case for melanoma recognition, where some black spots appear in the dermoscopic image. Our proposed framework is insufficient to segment dermocopic image as Fig. \ref{fig:7} (a) and recognize melanoma as Fig. \ref{fig:7} (b), because the both skin lesion cases are a minority of lesion types found in ISBI 2016 and ISBI 2017 datasets. 

\section{Conclusion}

In this paper, we propose to integrate dermatologists clinical knowledge into the learning process of a knowledge-aware deep framework for collaborative skin lesion segmentation and melanoma recognition. In particular, we propose a lesion-based pooling and shape extraction method to transfer the lesion structure information from the skin lesion segmentation into the melanoma recognition. It embeds the morphological operation within the deep network, which assists the network to learn more informative feature representation for melanoma recognition. We also propose a diagnosis guided feature fusion to pass the lesion class information from the melanoma recognition into the skin lesion segmentation, which enhances the performance of our network in detecting different types of skin lesions from dermoscopic images. Furthermore, a recursive mutual learning mechanism is designed to boost the joint leaning capability of the network on skin lesion segmentation and melanoma recognition. Experimental results have shown the effectiveness of the proposed approach on skin lesion segmentation and melanoma recognition, which achieves state-of-the-art performance on two publicly available skin lesion datasets. The demonstrated effectiveness of the proposed network on skin lesion diagnosis supports an advanced way of the deep learning in real clinic application for medical image analysis. Our future work will focus on closer collaboration with the clinical doctors to further enhance the computer-aided disease diagnosis of skin lesion.

\bibliography{Refskin}

\begin{thebibliography}{10}
\expandafter\ifx\csname url\endcsname\relax
  \def\url#1{\texttt{#1}}\fi
\expandafter\ifx\csname urlprefix\endcsname\relax\def\urlprefix{URL }\fi
\expandafter\ifx\csname href\endcsname\relax
  \def\href#1#2{#2} \def\path#1{#1}\fi

\bibitem{kawahara2018seven}
J.~Kawahara, S.~Daneshvar, G.~Argenziano, G.~Hamarneh, Seven-point checklist
  and skin lesion classification using multitask multimodal neural nets, IEEE
  J. Biomed. Health Informs. 23~(2) (2018) 538--546.

\bibitem{hu2018deep}
Z.~Hu, J.~Tang, Z.~Wang, K.~Zhang, L.~Zhang, Q.~Sun, Deep learning for
  image-based cancer detection and diagnosis- a survey, Pattern Recognit. 83
  (2018) 134--149.

\bibitem{yuan2017automatic}
Y.~Yuan, M.~Chao, Y.~Lo, Automatic skin lesion segmentation using deep fully
  convolutional networks with jaccard distance, IEEE Trans. Med. Imaging 36~(9)
  (2017) 1876--1886.

\bibitem{barata2020explainable}
C.~Barata, M.~Celebi, J.~Marques, Explainable skin lesion diagnosis using
  taxonomies, Pattern Recognit. (2020) 107413.

\bibitem{shuai2019toward}
B.~Shuai, H.~Ding, T.~Liu, G.~Wang, X.~Jiang, Toward achieving robust low-level
  and high-level scene parsing, IEEE Trans. Image Process. 28~(3) (2019)
  1378--1390.

\bibitem{ding2019semantic}
H.~Ding, X.~Jiang, B.~Shuai, A.~Q. Liu, G.~Wang, Semantic correlation promoted
  shape-variant context for segmentation, in: Proceedings of the IEEE/CVF
  Conference on Computer Vision and Pattern Recognition, 2019, pp. 8885--8894.

\bibitem{ding2020phraseclick}
H.~Ding, S.~Cohen, B.~Price, X.~Jiang, Phraseclick: toward achieving flexible
  interactive segmentation by phrase and click, in: European Conference on
  Computer Vision, Springer, 2020, pp. 417--435.

\bibitem{ding2021vision}
H.~Ding, C.~Liu, S.~Wang, X.~Jiang, Vision-language transformer and query
  generation for referring segmentation, in: Proceedings of the IEEE/CVF
  International Conference on Computer Vision, 2021, pp. 16321--16330.

\bibitem{ding2021interaction}
H.~Ding, H.~Zhang, J.~Liu, J.~Li, Z.~Feng, X.~Jiang, Interaction via
  bi-directional graph of semantic region affinity for scene parsing, in:
  Proceedings of the IEEE/CVF International Conference on Computer Vision,
  2021, pp. 15848--15858.

\bibitem{liu2021few}
W.~Liu, Z.~Wu, H.~Ding, F.~Liu, J.~Lin, G.~Lin, Few-shot segmentation with
  global and local contrastive learning, arXiv preprint arXiv:2108.05293.

\bibitem{liuweid2021few}
W.~Liu, C.~Zhang, H.~Ding, T.-Y. Hung, G.~Lin, Few-shot segmentation with
  optimal transport matching and message flow, arXiv preprint arXiv:2108.08518.

\bibitem{zhang2021prototypical}
H.~Zhang, H.~Ding, Prototypical matching and open set rejection for zero-shot
  semantic segmentation, in: Proceedings of the IEEE/CVF International
  Conference on Computer Vision, 2021, pp. 6974--6983.

\bibitem{zortea2017simple}
M.~Zortea, E.~Flores, J.~Scharcanski, A simple weighted thresholding method for
  the segmentation of pigmented skin lesions in macroscopic images, Pattern
  Recognit. 64 (2017) 92--104.

\bibitem{schmid1999segmentation}
P.~Schmid, Segmentation of digitized dermatoscopic images by two-dimensional
  color clustering, IEEE Trans. Med. Imaging 18~(2) (1999) 164--171.

\bibitem{zhou2009anisotropic}
H.~Zhou, G.~Schaefer, A.~Sadka, M.~Celebi, Anisotropic mean shift based fuzzy
  c-means segmentation of deroscopy images, IEEE J. Sel. Top. Signal Process.
  3~(1) (2009) 26--34.

\bibitem{zhou2011gradient}
H.~Zhou, G.~Schaefer, M.~Celebi, F.~Lin, T.~Liu, Gradient vector flow with mean
  shift for skin lesion segmentation, Comput. Med. Imag. Grap. 35~(2) (2011)
  121--127.

\bibitem{ma2016novel}
Z.~Ma, J.~M.~R. Tavares, A novel approach to segment skin lesions in
  dermoscopic images based on a deformable model, IEEE J. Biomed. Health Inform
  20~(2) (2016) 615--623.

\bibitem{bi2020multi}
L.~Bi, D.~Feng, M.~Fulham, J.~Kim, Multi-label classification of multi-modality
  skin lesion via hyper-connected convolutional neural network, Pattern
  Recognit. (2020) 107502.

\bibitem{sadeghi2013detection}
M.~Sadeghi, T.~Lee, D.~McLean, H.~Lui, M.~Atkins, Detection and analysis of
  irregular streaks in dermoscopic images of skin lesions, IEEE Trans. Biomed.
  Eng. 32~(5) (2013) 849--861.

\bibitem{saez2014model}
A.~Saez, C.~Serrano, B.~Acha, Model-based classification methods of global
  patterns in dermoscopic images, IEEE Trans. Med. Imaging 33~(5) (2014)
  1137--1147.

\bibitem{yu2018melanoma}
Z.~Yu, X.~Jiang, F.~Zhou, J.~Qin, D.~Ni, S.~Chen, B.~Lei, T.~Wang, Melanoma
  recognition in dermoscopy images via aggregated deep convolutional features,
  IEEE Trans. Biomed. Eng. 66~(4) (2018) 1006--1016.

\bibitem{yang2018classification}
J.~Yang, F.~Xie, H.~Fan, Z.~Jiang, J.~Liu, Classification for dermoscopy images
  using convolutional neural networks based on region average pooling, IEEE
  Access 6 (2018) 65130--65138.

\bibitem{cai2021unified}
Y.~Cai, Y.~Wang, Y.~Zhu, T.-J. Cham, J.~Cai, J.~Yuan, J.~Liu, C.~Zheng, S.~Yan,
  H.~Ding, et~al., A unified 3d human motion synthesis model via conditional
  variational auto-encoder, in: Proceedings of the IEEE/CVF International
  Conference on Computer Vision, 2021, pp. 11645--11655.

\bibitem{sun2021m2iosr}
X.~Sun, H.~Ding, C.~Zhang, G.~Lin, K.-V. Ling, M2iosr: Maximal mutual
  information open set recognition, arXiv preprint arXiv:2108.02373.

\bibitem{wang2021adaptive}
Y.~Wang, Y.~Cai, Y.~Liang, H.~Ding, C.~Wang, S.~Bhatia, B.~Hooi, Adaptive data
  augmentation on temporal graphs, Advances in Neural Information Processing
  Systems 34.

\bibitem{li2021else}
T.~Li, Q.~Ke, H.~Rahmani, R.~E. Ho, H.~Ding, J.~Liu, Else-net: Elastic semantic
  network for continual action recognition from skeleton data, in: Proceedings
  of the IEEE/CVF International Conference on Computer Vision, 2021, pp.
  13434--13443.

\bibitem{tong2021directed}
Z.~Tong, Y.~Liang, H.~Ding, Y.~Dai, X.~Li, C.~Wang, Directed graph contrastive
  learning, Advances in Neural Information Processing Systems 34.

\bibitem{li2021mine}
J.~Li, Z.~Feng, Q.~She, H.~Ding, C.~Wang, G.~H. Lee, Mine: Towards continuous
  depth mpi with nerf for novel view synthesis, in: Proceedings of the IEEE/CVF
  International Conference on Computer Vision, 2021, pp. 12578--12588.

\bibitem{mei2020object}
J.~Mei, H.~Ding, X.~Jiang, Object 6d pose estimation with non-local attention,
  in: Twelfth International Conference on Digital Image Processing (ICDIP
  2020), Vol. 11519, International Society for Optics and Photonics, 2020, p.
  115191H.

\bibitem{bi2017dermoscopic}
L.~Bi, J.~Kim, E.~Ahn, A.~Kumar, M.~Fulham, D.~Feng, Dermoscopic image
  segmentation via multi-stage fully convolutional networks, IEEE Trans.
  Biomed. Eng 64~(9) (2017) 2065--2074.

\bibitem{yuan2017improving}
Y.~Yuan, Y.~Lo, Improving dermoscopic image segmentation with enhanced
  convolutional-deconvolutional networks, IEEE J. Biomed. Health Inform 23~(2)
  (2017) 519--526.

\bibitem{wang2019dermoscopic}
X.~Wang, H.~Ding, X.~Jiang, Dermoscopic image segmentation through the enhanced
  high-level parsing and class weighted loss, in: Proceedings of the IEEE
  International Conference on Image Processing, 2019, pp. 245--249.

\bibitem{zhang2019dsm}
G.~Zhang, X.~Shen, S.~Chen, L.~Liang, Y.~Luo, J.~Yu, J.~Lu, Dsm: A deep
  supervised multi-scale network learning for skin cancer segmentation, IEEE
  Access 7 (2019) 140936--140945.

\bibitem{singh2019fca}
V.~Singh, M.~Abdel-Nasser, H.~Rashwan, F.~Akram, N.~Pandey, A.~Lalande,
  B.~Presles, S.~Romani, D.~Puig, Fca-net: Adversarial learning for skin lesion
  segmentation based on multi-scale features and factorized channel attention,
  IEEE Access 7 (2019) 130552--130565.

\bibitem{harangi2018skin}
B.~Harangi, Skin lesion classification with ensembles of deep convolutional
  neural networks, J. Biomed. Inform. 86 (2018) 25--32.

\bibitem{mahbod2019fusing}
A.~Mahbod, G.~Schaefer, I.~Ellinger, R.~Ecker, A.~Pitiot, C.~Wang, Fusing
  fine-tuned deep features for skin lesion classification, Comput. Med. Imaging
  Graph. 71 (2019) 19--29.

\bibitem{liang2019multi}
R.~Liang, Q.~Wu, X.~Yang, Multi-pooling attention learning for melanoma
  recognition, in: 2019 Digital Image Computing: Techniques and Applications
  (DICTA), IEEE, 2019, pp. 1--6.

\bibitem{gessert2019skin}
N.~Gessert, T.~Sentker, F.~Madesta, R.~Schmitz, H.~Kniep, I.~Baltruschat,
  R.~Werner, A.~Schlaefer, Skin lesion classification using cnns with
  patch-based attention and diagnosis-guided loss weighting, IEEE Trans.
  Biomed. Eng.

\bibitem{sultana2018deep}
N.~Sultana, B.~Mandal, N.~Puhan, Deep residual network with regularised fisher
  framework for detection of melanoma, IET Computer Vision 12~(8) (2018)
  1096--1104.

\bibitem{hagerty2019deep}
J.~Hagerty, J.~Stanley, H.~Almubarak, N.~Lama, R.~Kasmi, P.~Guo, R.~Drugge,
  H.~Rabinovitz, M.~Olivero, W.~Stoecker, Deep learning and handcrafted method
  fusion: Higher diagnostic accuracy for melanoma dermoscopy images, IEEE J.
  Biomed. Health Informs. 23~(4) (2019) 1385--1391.

\bibitem{zhang2019attention}
J.~Zhang, Y.~Xie, Y.~Xia, C.~Shen, Attention residual learning for skin lesion
  classification, IEEE Trans. Med. Imaging 38~(9) (2019) 2092--2103.

\bibitem{yu2017automated}
L.~Yu, H.~Chen, Q.~Dou, J.~Qin, P.~Heng, Automated melanoma recognition in
  dermoscopy images via very deep residual networks, IEEE Trans. Med. Imaging
  36~(4) (2017) 994--1004.

\bibitem{he2016deep}
K.~He, X.~Zhang, S.~Ren, J.~Sun, Deep residual learning for image recognition,
  in: Proceedings of the IEEE conference on computer vision and pattern
  recognition, 2016, pp. 770--778.

\bibitem{he2015spatial}
K.~He, X.~Zhang, S.~Ren, J.~Sun, Spatial pyramid pooling in deep convolutional
  networks for visual recognition, IEEE Trans. Pattern Anal. Mach.Intell.
  37~(9) (2015) 1904--1916.

\bibitem{gonzalez2018dermaknet}
I.~Gonz{\'a}lez-D{\'\i}az, Dermaknet: Incorporating the knowledge of
  dermatologists to convolutional neural networks for skin lesion diagnosis,
  IEEE J. Biomed. Health Informs. 23~(2) (2018) 547--559.

\bibitem{xie2016melanoma}
F.~Xie, H.~Fan, Y.~Li, Z.~Jiang, R.~Meng, A.~Bovik, Melanoma classification on
  dermoscopy images using a neural network ensemble model, IEEE Trans. Med.
  Imaging 36~(3) (2016) 849--858.

\bibitem{tajeddin2018melanoma}
N.~Tajeddin, B.~Asl, Melanoma recognition in dermoscopy images using lesion's
  peripheral region information, Comput. Meth. Prog. Bio. 163 (2018) 143--153.

\bibitem{riaz2018active}
F.~Riaz, S.~Naeem, R.~Nawaz, M.~Coimbra, Active contours based segmentation and
  lesion periphery analysis for characterization of skin lesions in dermoscopy
  images, IEEE J. Biomed. Health Informs. 23~(2) (2018) 489--500.

\bibitem{liu2021towards}
C.~Liu, H.~Ding, X.~Jiang, Towards enhancing fine-grained details for image
  matting, in: Proceedings of the IEEE/CVF Winter Conference on Applications of
  Computer Vision, 2021, pp. 385--393.

\bibitem{mei2019deepdeblur}
J.~Mei, Z.~Wu, X.~Chen, Y.~Qiao, H.~Ding, X.~Jiang, Deepdeblur: text image
  recovery from blur to sharp, Multimedia tools and applications 78~(13) (2019)
  18869--18885.

\bibitem{liu2019feature}
J.~Liu, H.~Ding, A.~Shahroudy, L.-Y. Duan, X.~Jiang, G.~Wang, A.~C. Kot,
  Feature boosting network for 3d pose estimation, IEEE transactions on pattern
  analysis and machine intelligence 42~(2) (2019) 494--501.

\bibitem{al2018skin}
M.~Al-masni, M.~Al-antari, M.~Choi, S.~Han, T.~Kim, Skin lesion segmentation in
  dermoscopy images via deep full resolution convolutional networks, Comput.
  Meth. Prog. Bio. 162 (2018) 221--231.

\bibitem{li2018dense}
H.~Li, X.~He, F.~Zhou, Z.~Yu, D.~Ni, S.~Chen, T.~Wang, B.~Lei, Dense
  deconvolutional network for skin lesion segmentation, IEEE J. Biomed. Health
  Inform 23~(2) (2018) 527--537.

\bibitem{tu2019dense}
W.~Tu, X.~Liu, W.~Hu, Z.~Pan, Dense-residual network with adversarial learning
  for skin lesion segmentation, IEEE Access 7 (2019) 77037--77051.

\bibitem{bi2019step}
L.~Bi, J.~Kim, E.~Ahn, A.~Kumar, D.~Feng, M.~Fulham, Step-wise integration of
  deep class-specific learning for dermoscopic image segmentation, Pattern
  Recognit. 85 (2019) 78--89.

\bibitem{tang2019efficient}
P.~Tang, Q.~Liang, X.~Yan, S.~Xiang, W.~Sun, D.~Zhang, G.~Coppola, Efficient
  skin lesion segmentation using separable-unet with stochastic weight
  averaging, Comput. Meth. Prog. Bio. 178 (2019) 289--301.

\bibitem{nasr2019dense}
E.~Nasr-Esfahani, S.~Rafiei, M.~Jafari, N.~Karimi, J.~Wrobel, S.~Samavi,
  S.~Soroushmehr, Dense pooling layers in fully convolutional network for skin
  lesion segmentation, Comput. Med. Imaging Graph. 78 (2019) 101658.

\bibitem{wei2019attention}
Z.~Wei, H.~Song, L.~Chen, Q.~Li, G.~Han, Attention-based denseunet network with
  adversarial training for skin lesion segmentation, IEEE Access 7 (2019)
  136616--136629.

\bibitem{wang2019bi}
X.~Wang, X.~Jiang, H.~Ding, J.~Liu, Bi-directional dermoscopic feature learning
  and multi-scale consistent decision fusion for skin lesion segmentation, IEEE
  Trans. Image Process. 29 (2020) 3039--3051.

\bibitem{esteva2017dermatologist}
A.~Esteva, B.~Kuprel, R.~Novoa, J.~Ko, S.~Swetter, H.~Blau, S.~Thrun,
  Dermatologist-level classification of skin cancer with deep neural networks,
  Nature 542~(7639) (2017) 115.

\bibitem{gu2019progressive}
Y.~Gu, Z.~Ge, C.~Bonnington, J.~Zhou, Progressive transfer learning and
  adversarial domain adaptation for cross-domain skin disease classification,
  IEEE J. Biomed. Health Informs. 24~(5) (2019) 1379--1393.

\bibitem{al2020multiple}
M.~Al-masni, D.~Kim, T.~Kim, Multiple skin lesions diagnostics via integrated
  deep convolutional networks for segmentation and classification, Comput.
  Meth. Prog. Bio. (2020) 105351.

\bibitem{xie2020mutual}
Y.~Xie, J.~Zhang, Y.~Xia, C.~Shen, A mutual bootstrapping model for automated
  skin lesion segmentation and classification, IEEE Trans. Med. Imaging 39~(7)
  (2020) 2482--2493.

\bibitem{yu2018recurrent}
Q.~Yu, L.~Xie, Y.~Wang, Y.~Zhou, E.~Fishman, A.~Yuille, Recurrent saliency
  transformation network: Incorporating multi-stage visual cues for small organ
  segmentation, in: Proceedings of the IEEE Conference on Computer Vision and
  Pattern Recognition, 2018, pp. 8280--8289.

\bibitem{zhao2019recursive}
S.~Zhao, Y.~Dong, E.~Chang, Y.~Xu, Recursive cascaded networks for unsupervised
  medical image registration, in: Proceedings of the IEEE International
  Conference on Computer Vision, 2019, pp. 10600--10610.

\bibitem{timmurphy16}
I.~2016, Skin lesion analysis towards melanoma detection,
  \url{https://challenge.kitware.com/\#phase/566744dccad3a56fac786787.},
  [Online].

\bibitem{timmurphy17}
I.~2017, Skin lesion analysis towards melanoma detection,
  \url{https://challenge.kitware.com/#phase/584b0afacad3a51cc66c8e24.},
  [Online].

\bibitem{russakovsky2015imagenet}
O.~Russakovsky, J.~Deng, H.~Su, J.~Krause, S.~Satheesh, S.~Ma, Z.~Huang,
  A.~Karpathy, A.~Khosla, M.~Bernstein, et~al., Imagenet large scale visual
  recognition challenge, Int. J. Comput. Vision 115~(3) (2015) 211--252.

\bibitem{ding2018context}
H.~Ding, X.~Jiang, B.~Shuai, A.~Liu, G.~Wang, Context contrasted feature and
  gated multi-scale aggregation for scene segmentation, in: Proceedings of the
  IEEE Conference on Computer Vision and Pattern Recognition, 2018, pp.
  2393--2402.

\bibitem{long2015fully}
J.~Long, E.~Shelhamer, T.~Darrell, Fully convolutional networks for semantic
  segmentation, in: Proceedings of the IEEE Conference on Computer Vision and
  Pattern Recognition, 2015, pp. 3431--3440.

\bibitem{ding2020semantic}
H.~Ding, X.~Jiang, B.~Shuai, A.~Liu, G.~Wang, Semantic segmentation with
  context encoding and multi-path decoding, IEEE Trans. Image Process. 29
  (2020) 3520--3533.

\bibitem{ding2019boundary}
H.~Ding, X.~Jiang, A.~Liu, N.~Thalmann, G.~Wang, Boundary-aware feature
  propagation for scene segmentation, in: Proceedings of the IEEE International
  Conference on Computer Vision, 2019, pp. 6819--6829.

\bibitem{selvaraju2017grad}
R.~Selvaraju, M.~Cogswell, A.~Das, R.~Vedantam, D.~Parikh, D.~Batra, Grad-cam:
  Visual explanations from deep networks via gradient-based localization, in:
  Proceedings of the IEEE international conference on computer vision, 2017,
  pp. 618--626.

\end{thebibliography}

\end{document}